\documentclass[preprint,pra,aps,showkeys,showpacs,superscriptaddress]{revtex4-1}

\usepackage{ucs}
\usepackage[english]{babel}
\usepackage{amsmath}
\usepackage{amssymb,amsfonts}
\usepackage{color}

\usepackage{graphicx}
\usepackage{dcolumn}
\usepackage{bm}
\usepackage[utf8]{inputenc}
\usepackage[T1]{fontenc}
\usepackage{tikz}
\usetikzlibrary{shapes.misc}
\usepackage{mathrsfs}
\usepackage[T1]{fontenc}
\usepackage{scrpage2}

\begin{document}

\title{Intensified antibunching via feedback-induced quantum interference}

\author{Yiping Lu}
\affiliation{Center for quantum technology research, School of physics, Beijing Institute of Technology, Haidian District, Beijing 100081, P.R. China}%
\affiliation{Technische Universit\"at Berlin, Institut f\"ur Theoretische Physik, Nichtlineare Optik und Quantenelektronik, Hardenbergstraße 36, 10623 Berlin, Germany}
\author{Nicolas L. Naumann}%
\affiliation{Technische Universit\"at Berlin, Institut f\"ur Theoretische Physik, Nichtlineare Optik und Quantenelektronik, Hardenbergstraße 36, 10623 Berlin, Germany}
\author{Javier Cerrillo}
\affiliation{Institut f\"ur Theoretische Physik, Technische Universit\"at Berlin, Hardenbergstr. 36, 10623 Berlin, Germany}
\author{Qing Zhao}
\affiliation{Center for quantum technology research, School of physics, Beijing Institute of Technology, Haidian District, Beijing 100081, P.R. China}
\author{Andreas Knorr}
\affiliation{Technische Universit\"at Berlin, Institut f\"ur Theoretische Physik, Nichtlineare Optik und Quantenelektronik, Hardenbergstraße 36, 10623 Berlin, Germany}
\author{Alexander Carmele}
\email{alex@itp.tu-berlin.de}
\affiliation{Technische Universit\"at Berlin, Institut f\"ur Theoretische Physik, Nichtlineare Optik und Quantenelektronik, Hardenbergstraße 36, 10623 Berlin, Germany}

\date{\today}

\begin{abstract}
We numerically show that time delayed coherent feedback controls the 
statistical output characteristics of driven quantum emitters. 
Quantum feedback allows to enhance or suppress a wide range of classical and 
nonclassical features of the emitted quantum light. 
As exemplary quantum system, we use a pumped cavity containing two emitters. 
By applying phase-selective feedback, we demonstrate that photon antibunching
and bunching can be increased in orders of magnitude due to intrinsically 
and externally controllabe quantum interferences. 
Our modelling is based on a fully non-Markovian quantum simulation of a 
structured photon continuum. 
We show that an approximative method in the Schr\"odinger picture allows a 
very good estimate for quantum feedback induced features for low pump rates.
\end{abstract}

\pacs{42.50.Ar, 42.50.Pq, 42.50.Ct, 02.30.Ks}
\maketitle

\section{Introduction\label{sec:level1}}

In the pursuit of exploiting quantum systems for technological applications
many challenges have to be overcome. One of those is the
limited coherence time of quantum systems, during which quantum
correlations are dominant in the system \cite{epjd1,wiseman1}.
To stabilize quantum features, different mechanisms have been proposed
such as gate purification \cite{prl2}, error correction \cite{pra3}, and feedback \cite{wiseman1}.
Using quantum feedback extrinsically, a stabilization of certain states
has been achieved successfully \cite{prl4}.

Motivated by classical Pyragas control \cite{pla5,sfb910},
recent experiments start to investigate the role of a non-negligible
delay time close to the quantum regime \cite{nc6,1367-2630-15-2-025030,hoi_probing_2015}.

In the regime of classical optics, time-delayed self-feedback was found
to have a significant impact on
the dynamics of a semiconductor laser \cite{jqe6}.
This has been exploited to control laser systems \cite{pre7}.
A more recent proposal uses feedback as a mechanism to reduce the jitter
in pulsed lasers or the optoelectronic control of the device based on a 
classical series of measurements \cite{njp8,2017_03_Munnelly}.
Other possible applications include optomechanical setups 
\cite{PhysRevX.7.011001}.
However, for time critical applications, coherent control may have advantages over
measurement based schemes \cite{1367-2630-16-7-073036}.

In the classical regime, time delayed feedback is usually modeled
using delayed differential
equations \cite{jqe6}.
However, in the quantum regime a model for the description of
nonclassical light is necessary, where the electronic and photonic degrees
of freedom are treated on an equal footing. 
Here, the feedback channel is understood as a reservoir of infinite
modes, which induces the
delay time by its structured nature, corresponding to a non Markovian
bath \cite{UDornerPZoller,PhysRevA.10.1096}.
In contrast to other models for structured baths
\cite{RevModPhys.88.021002,*PhysRevLett.82.1801,*0034-4885-77-9-094001}, in this case
one specific delay time characterizes the reservoir response.

In the limit of high photon number, semiclassical approximations allow to
derive manageable equations
and establish the ties to the classical regime \cite{pra11,*pra12,josa19}.

However, an alternate approach in recent experiments aims at reaching
the quantum optical regime \cite{albert_observing_2011}, so that a
semiclassical description could become insufficient.
In this regime of few excitations,
time delayed quantum optical feedback has been proposed e.g. to
stabilize Rabi oscillations
\cite{prl13} and to control entanglement
\cite{prl14,*pra15}.
However, in the theoretical description of these proposals, initial states 
with a fixed number of excitations have been assumed in order to close 
the respective set of equations of motions. 
To examine the photon statistics of a driven optical system, the number of 
excitations are initially not limited and new methods are needed to describe
quantum feedback effects in driven-damped setups.

While there are special cases where driven quantum systems can be considered
analytically \cite{pra16,*PhysRevA.94.023809},
for most cases a numerical solution is necessary. To solve this problem
methods have been proposed employing
the Liouville space \cite{prl17}, Matrix product state (MPS) evolution  \cite{prl18},
and the Heisenberg picture \cite{josa19}.

In this manuscript, we propose an alternative, approximative method in
the Schr\"odinger picture, which is applicable in the
regime of small pump rates.
This method is easy to implement, since it basically involves the numerical
evaluation of Schr\"odinger's equation with an effective, dissipative
Hamiltonian \cite{ScieRe,quantumnoise}.
This model agrees well with numerically exact models in the low pumping regime
and allows a very good estimate of quantum feedback effects in open, driven
cavity QED systems.

We show theoretically how time delayed feedback may be used to control
classical and non classical properties of a quantum system. 
Thus, feedback
can be used to amplify bunching and antibunching, as well as to increase
the amount of entanglement between two two level systems inside a cavity.
This will be instantiated with a system consisting of two emitters described 
as two level systems inside a cavity, exchanging quantum excitations only 
via the cavity mode.
This system was recently proposed to show a connection between the photon statistics 
of the output field and the entanglement of the two level systems inside the
cavity \cite{ScieRe}. 
This connection is useful to have a first grasp at possible
quantum correlations inside a system, as traditional ways to gain information 
about the correlations inside a system are more involved. 
One may try to determine the fidelity for a certain state or even consider 
reconstructing the full density matrix \cite{Lu2015161,*lu_minimum_2016}, which contains all information about a 
system \cite{PhysRevA.64.052312,*PhysRevA.66.012303}.

The manuscript is structured as follows: In Sec. \ref{sec:model}
we will introduce the approximative model to describe feedback in
the low pumping regime in the system consisting of two emitters
inside a cavity subject to feedback. In the Sec. \ref{sec:control},
we will show how time delayed feedback can be used to control the
photon statistics of the emitted light field and emphasize classical
as well as nonclassical photon statistics. In Sec. \ref{sec:entanglement},
we will discuss the relation between the entanglement in the emitters to
the photon statistics when applying feedback to the system.
A summary concludes our manuscript.

\section{Model\label{sec:model}}

\begin{figure}[!h]
\centering
\scalebox{0.45}{
\begin{tikzpicture}

\fill[blue!40] (0,0) -- (8,0) -- (8,2) -- (0,2) -- cycle;
\draw (0,0) -- (8,0) -- (8,2) -- (0,2) -- cycle;

\draw (8,3) -- (8,-1);

\draw (8,2) -- (8.5,1.5);
\begin{scope}[shift = {(0,-0.25)}]
\draw (8,2) -- (8.5,1.5);
\end{scope}
\begin{scope}[shift = {(0,-0.5)}]
\draw (8,2) -- (8.5,1.5);
\end{scope}
\begin{scope}[shift = {(0,-0.75)}]
\draw (8,2) -- (8.5,1.5);
\end{scope}
\begin{scope}[shift = {(0,-1)}]
\draw (8,2) -- (8.5,1.5);
\end{scope}
\begin{scope}[shift = {(0,-1.25)}]
\draw (8,2) -- (8.5,1.5);
\end{scope}
\begin{scope}[shift = {(0,-1.5)}]
\draw (8,2) -- (8.5,1.5);
\end{scope}
\begin{scope}[shift = {(0,-1.75)}]
\draw (8,2) -- (8.5,1.5);
\end{scope}
\begin{scope}[shift = {(0,-2)}]
\draw (8,2) -- (8.5,1.5);
\end{scope}
\begin{scope}[shift = {(0,-2.25)}]
\draw (8,2) -- (8.5,1.5);
\end{scope}
\begin{scope}[shift = {(0,-2.5)}]
\draw (8,2) -- (8.5,1.5);
\end{scope}
\begin{scope}[shift = {(0,-2.75)}]
\draw (8,2) -- (8.5,1.5);
\end{scope}
\begin{scope}[shift = {(0,.25)}]
\draw (8,2) -- (8.5,1.5);
\end{scope}
\begin{scope}[shift = {(0,0.5)}]
\draw (8,2) -- (8.5,1.5);
\end{scope}
\begin{scope}[shift = {(0,0.75)}]
\draw (8,2) -- (8.5,1.5);
\end{scope}

\draw (8,-2) node[scale = 1.5]{Mirror};

\draw[line width=2.5pt, dotted, <->] (1,1) -- (8,1);
\draw [blue!40,fill = blue!40](4,0.5) rectangle (5,1.5);
\draw (4.5,1) node{\huge{$L$}};
\draw (4.5,2.6) node{\huge{$d_k,d_k^\dagger$}};

\draw[very thick, dashed] (1,1) -- (-0.75,-2.);
\draw[very thick, dashed] (1,1) -- (3.5,-2);

\begin{scope}[shift = {(-0.5,-5)}]
\draw[very thick] (0,0) to[bend left=20] (0,3) ;
\draw[very thick] (3.75,0) to[bend right=20] (3.75,3) ;
\draw[line width=2pt] (0,1) -- (1,1);
\draw[line width=2pt] (2,1) -- (3,1);
\draw[line width=2pt] (0,2) -- (1,2);
\draw[line width=2pt] (2,2) -- (3,2);
\draw[dashed,very thick,->] (0.5,2) -- (0.5,1) node[midway,right,scale =
1.5]{$\gamma$};
\draw[dashed,very thick,->] (2.5,2) -- (2.5,1) node[midway,right,scale =
1.5]{$\gamma$};
\draw (1.4,2) node[scale=1.5]{$|e\rangle$};
\draw (1.4,1) node[scale=1.5]{$|g\rangle$};
\draw (3.4,2) node[scale=1.5]{$|e\rangle$};
\draw (3.4,1) node[scale=1.5]{$|g\rangle$};
\end{scope}

\begin{scope}[shift = {(-3.5,-3.5)}]
\draw (0,0) ellipse (0.25cm and 0.5cm);
\draw (0,0.5) -- (2,0.5);
\draw (0,-0.5) -- (2,-0.5);
\draw (2,0.5) to[bend left=50] (2,-0.5) ;
\draw (1,0.5) node[above,scale = 2]{$\epsilon$};
\end{scope}

\begin{scope}[shift = {(+3.5,-3.5)}]
\draw (1,-0.5) node[above,scale = 2]{$c,c^{\dagger}$};
\end{scope}
\end{tikzpicture}
}
\caption{Realization of quantum feedback via photonic waveguide. Two two-level emitters trapped in a small cavity interact with the photons inside the cavity. The driving field $\epsilon$ is weakly coupled to the cavity mode with frequency $w_L$. The cavity losses are fed back into the cavity from the waveguide with the distance $L$ from cavity. Furthermore, the two level systems may decay spontaneously with the decay rate $\gamma$.}\label{fig:sketch2map}
\end{figure}
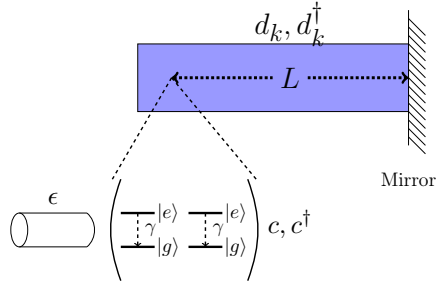

To investigate the system dynamics, we start from a model Hamiltonian describing the system depicted in Fig. \ref{fig:sketch2map}. Here, a small pumped cavity contains two two level systems, which is coupled to an external feedback channel. The frequency of the external pump laser may be detuned from the resonance of the cavity and the two level systems.
Such systems can be realized, e.g. by nanoscale optical positioning in 
integrated quantum photonic devices. \cite{sapienza_nanoscale_2015}

We start from the Hamiltonian in the frame rotating with the external driving frequency \cite{ScieRe,JuliaPRA} ($\hbar=1$) 
\begin{align}
H&=\Delta a^\dag a+ \sum_{i=1}^{2}\delta\sigma_i^+\sigma_i^-+ \sum_{i=1}^{2}g_i(\sigma_i^+a+a^\dag\sigma_i^-)\nonumber\\
&+\epsilon(a^\dag+a)+\int dk(G(k,t)a^\dag d_k+G^*(k,t)d_k^\dag a)\nonumber\\
&-\frac{i}{2}\gamma\sum_{i=1}^{2}\sigma_i^+\sigma_i^-.
\label{eq:H}
\end{align}
Here, $a$ ($a^\dag$) is the annihilation (creation) operator of the photon in the cavity, $\sigma_i^+$ is the raising operator, $\sigma_i^-$ is the lowering operator of the $i$-th two level system and $\sigma_i^-=|g_i\rangle\langle e_i|$. Here, $|e_i\rangle$ is excited and $|g_i\rangle$ the ground state of the $i$th two-level system. The cavity is pumped by an external driving with the strength $\epsilon$ and the frequency $\omega_L$. The detuning from the cavity mode with frequency $\omega_0$ is $\Delta=\omega_0-\omega_L$. The detuning from the resonance is $\delta=\omega_{\mathrm{eg}}-\omega_L$, where $\omega_{\mathrm{eg}}$ is the frequency of the resonance in the two level system. The $i$-th two level system is coupled to the cavity mode with $g_i$. The spontaneous emission of the emitters is described approximately by a non-hermitian part in the Hamiltonian, where the phenomenological decay rate into not considered modes is given as $\gamma$. We assume that the cavity is resonant with the emitters, i.e. $\delta=\Delta$, and the coupling is the same for all emitters $g_1=g_2=g$.
Feedback is described by an external continuum of modes, where $d_k$ ($d_k^\dag$) is the annihilation (creation) operator of a photon in the continuum with 1D wave vector $k$. The cavity mode is coupled to the external mode with
\begin{equation}
G(k,t)=G_0\sin(kL)\exp[i(\omega_L-\omega_k)t].
\end{equation}
Here $\omega_k=c|k|$ with $c$ being the speed of light in the waveguide. $G_0$ is the bare tunnel coupling strength between cavity photons and external modes in continuum. The frequencies $w_L$ and $w_k$ correspond to the external driving frequency and the half-cavity modes with wave number $k$, respectively. We assume that the feedback channel is characterized by a single time delay, $\tau=2L/c$. It is verified that the two-photon limit is sufficient to describe the system when the driving field $\epsilon$ is weak by comparison with the MPS evolution method \cite{prl18} at the end of this section. Therefore, the wave function of the state of the entire system in this case can be written as

\begin{align}
\nonumber
|\varphi(t)\rangle &=\int dk\int dk'C_{gg0kk'}|g,g,0,\{k\},\{k'\}\rangle\\
\nonumber
&+\int dk C_{eg0k}|e,g,0,\{k\}\rangle+\int dk C_{gg0k}|g,g,0,\{k\}\rangle\\
\nonumber
&+\int dk C_{gg1k}|g,g,1,\{k\}\rangle+\int dk C_{ge0k}|g,e,0,\{k\}\rangle\\
\nonumber
&+C_{ge10}|g,e,1,\{0\}\rangle+ C_{eg10}|e,g,1,\{0\}\rangle\\
\nonumber
&+C_{gg10}|g,g,1,\{0\}\rangle+C_{gg20}|g,g,2,\{0\}\rangle\\
\nonumber
&+C_{ee00}|e,e,0,\{0\}\rangle+C_{eg00}|e,g,0,\{0\}\rangle\\
&+C_{ge00}|g,e,0, \lbrace 0\rbrace\rangle+C_{gg00}|g,g,0,\lbrace 0\rbrace\rangle.
\label{eq:psi}
\end{align}

Here, we choose the basis $|i_1,i_2,i_{\mathrm{photon}},{k},{k'}\rangle$, in which $i_1$ and $i_2$ describe the two emitters, $i_{\mathrm{photon}}$ is the number of photons in the cavity mode, and $k$, $k'$ is the external mode in which a photon is present.
Correspondingly, the coefficients $C_{i_1,i_2,i_{\mathrm{photon}},k,k'}$ indicate the probability amplitudes of the state. For brevity, we suppressed the time dependence of the state coefficients. The equation of motion for the coefficients are derived from the Schr\"odinger equation and are given in the App. \ref{sec:eom}.

To investigate the photon statistics, we consider the second order correlation function \cite{Wall}, which reads in the Heisenberg picture
\begin{equation}
g^{(2)}(t,t')=\frac{\langle a^{\dagger}(t) a^{\dagger}(t+t') a(t+t') a(t)\rangle}{\langle a^{\dagger}(t) a(t)\rangle^2}.
\label{g2H}
\end{equation}
We will consider the case with $t'=0$, which becomes in the Schr\"odinger picture
\begin{equation}
g^{(2)}(t,0)=\frac{\langle\varphi(t)| a^{\dagger} a^{\dagger} a a|\varphi(t)\rangle}{\langle\varphi(t)| a^{\dagger} a|\varphi(t)\rangle^2}.
\label{g2S}
\end{equation}
Here, $\langle\varphi(t)| a^{\dagger} a|\varphi(t)\rangle$ is the mean intra-cavity photon number.
The second order correlation function characterizes the light field by putting the intensity in relation to the multi-photon part of the light field. A coherent field, as emitted by a laser, exhibits a value of one, showing no correlation between measuring two photons. When a value greater than one is observed, the probability of measuring two photons simultaneously is increased. A value of smaller than one is an indicator for a nonclassical state. This regime is called antibunching, since the probability to measure two photons simultaneously is decreased. The explicit expression for the correlation function in the present case is given in App. \ref{sec:eom}.
%
%
%

Furthermore, the concurrence is evaluated to quantify the entanglement between the two emitters inside the cavity. Concurrence serves as a measure for entanglement. Its maximal value is unity, indicating maximal entanglement, while the minimal value zero signifies a fully separable product state with no entanglement. The concurrence is defined as \cite{Woot}
\begin{equation}
C(\rho_{AB})=\max\{0, \sqrt{\lambda_1}-\sqrt{\lambda_2}-\sqrt{\lambda_3}-\sqrt{\lambda_4}\},
\end{equation}
where $\lambda_i$ is the square root of the $i$-th eigenvalue of the matrix $\rho_{AB}(\sigma_y\otimes\sigma_y)\rho_{AB}^*(\sigma_y\otimes\sigma_y)$ in decreasing order. 

Here, $\rho_{AB}$ is the reduced density matrix of the subsystem consisting of the two emitters, i.e.
\begin{align}
\rho_{AB}&=\mathrm{Tr}_{\text{photonic modes}}(\rho),
\end{align}
where $\rho=|\varphi(t)\rangle\langle\varphi(t)|$ is the density matrix for whole system.
The reduced density matrix is multiplied by the Tensor product of the Pauli matrices in the subspace of each emitter.

\subsection*{Comparison with MPS evolution method}

\begin{figure}[h]
\centering
\includegraphics[width=0.35\linewidth]{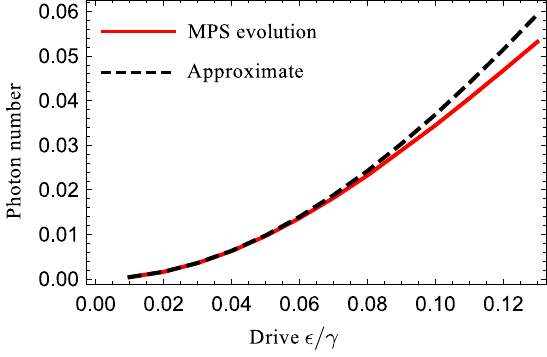}
\caption{Comparison of numerically exact MPS evolution method and the proposed approximative scheme. For low pump strengths the results coincide. Parameters: $\omega_0=1.1\times10^5\gamma$, $g=40\gamma$, and $G_0=\sqrt{\frac{2c\gamma}{\pi}}$, $\tau=0.005\gamma$.}
\label{fig:peps}
\end{figure}

To give the regime of validity of the approximation employed in this paper, we show in Fig. \ref{fig:peps} the photon number as computed by the approximative method used in this paper and the numerically exact MPS evolution method as proposed in Refs. \cite{prl18,PhysRevLett.95.110503,*PhysRevA.75.032311}, which relies on techniques developed in the context of condensed matter systems \cite{Schollwoeck201196,*PhysRevLett.91.147902}. In this paper we stay in the regime of low pumping, where the approximation is valid.

\section{Control of photon statistics \label{sec:control}}
\begin{figure}
\includegraphics[width=0.5\columnwidth]{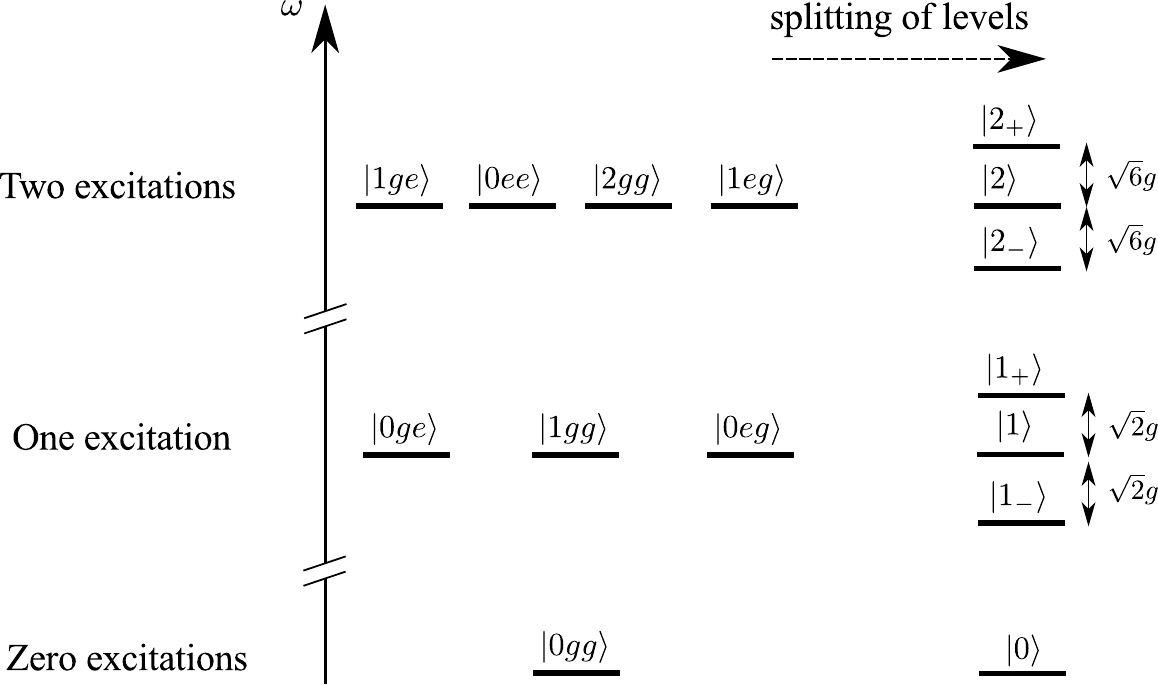}
\caption{ \label{fig:levels} Energy levels of the Hamiltonian \eqref{eq:H} without losses and feedback. On the left hand side, the basis states are shown, which correspond to the case without interaction. The interaction between the cavity and the two level systems lifts the degeneracy, so that specific transitions can be addressed by adjusting the frequency of the external optical pump. The definition of the eigenstates of the Hamiltonian with interaction (right hand side) can be found in App. \ref{sec:states}.}
\end{figure}

In order to investigate the impact of feedback on the photon statistics, we consider two scenarios in detail. In Fig. \ref{fig:levels}, we show the energy levels of the Hamiltonian without loss and feedback.
The interaction between the cavity mode and the emitters lifts the degeneracy. In App. \ref{sec:states}, we give the definition of the eigenstates of the Hamiltonian with interaction. These can be used to explain that certain transitions can be addressed by adjusting the driving laser frequency, so that specific states can be populated. We will investigate the transitions from the ground state to the single excitation state $|1_+\rangle$ and from the ground state to the two excitation state $|2_+\rangle$.

Now, we briefly discuss the case without feedback as discussed by Zhang et al. to establish a connection between the photon statistics and entanglement \cite{ScieRe}.
When considering the behavior of the system as a function of the detuning, there are two detunings of interest. Choosing $\Delta=\sqrt{2}g$, the single excitation state $|1_+\rangle$ is addressed in particular, since this is resonant via a single photon process with the energy of this state, cf. Fig. \ref{fig:levels}. The field emitted from the cavity in the case without feedback exhibits antibunching, since the superposed $|1_+\rangle$-state contains exactly one excitation. In contrast to this, pumping the two excitation state $|2_+\rangle$ via a two photon process by choosing $\Delta=\frac{\sqrt{6}}{2}g$, the emitted field shows bunching. In this case the state $|2_+\rangle$ contains exactly two excitations. These situations also exhibit a small degree of entanglement. By pumping weakly with a coherent field, the states cannot be prepared as pure states, since the major contribution to the state contains no excitation. Nonetheless, the weakly addressed states imprint the discussed signatures on the second order correlation function and the concurrence due to their properties.

In the following, we compare the system with feedback to a system without feedback.
For the reference system without feedback, the cavity loss is chosen such that the dynamics until $\tau$ is equal to the feedback case. We choose this, since we only want to consider the effects due to the self interference. This is achieved by setting $G_{\mathrm{no feedback}}(k,t)=G_0 \frac{1}{\sqrt{2}} \exp\left[ i(\omega_0-\omega_k) t \right]$.

In the following, we will consider short time delays, where $\tau\leq \frac{1}{\gamma}$, so that the system without feedback would not yet be in a stationary state.

\subsection*{Enhancement of antibunching}

\begin{figure}[!htb]
\includegraphics[width=0.5\columnwidth]{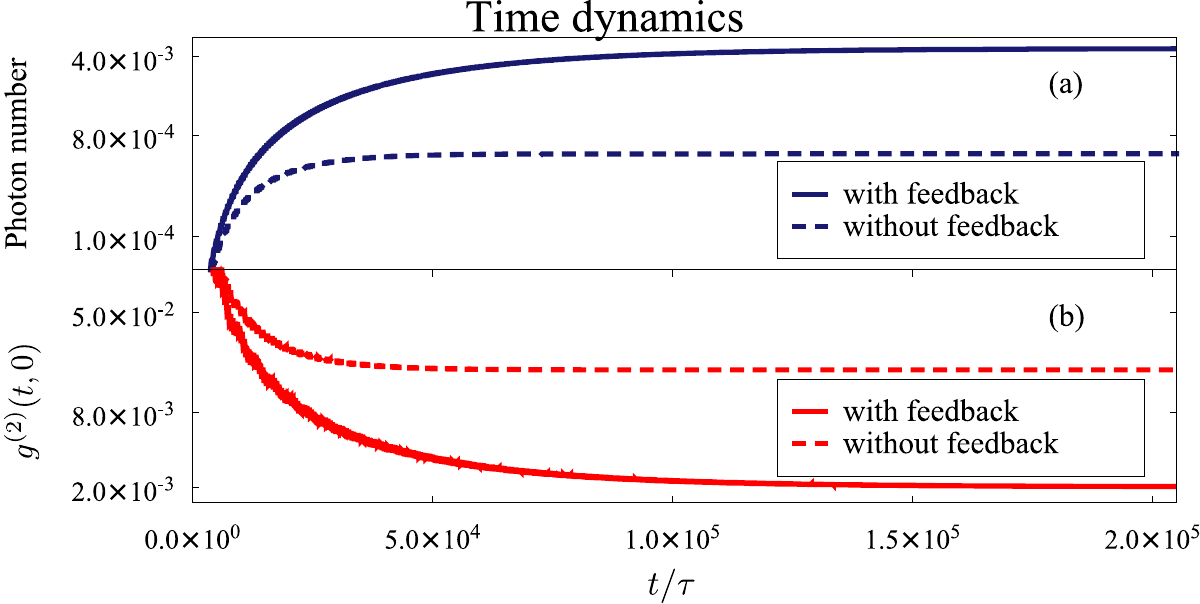}~~~%
	\caption{The time dynamics of (a) the photon number and (b) the $g^{(2)}(t,0)$ function for the case of photon antibunching with and without feedback. The dashed curves denote the case without feedback. The solid curves represent the case with feedback. We take $\Delta=\delta=\sqrt{2}g$, $\tau=4\pi/\omega_0$, $\omega_0=1.1\times10^5\gamma$, $G_0=\sqrt{2c\gamma/\pi},\epsilon=0.035\gamma$, and $g=40\gamma$.\label{fig:antibunching}}
\end{figure}

First, we consider the situation where the non-feedback case exhibits antibunching. This is achieved by choosing the detuning $\Delta=\sqrt{2} g$. Then, the single excitation state is pumped (cf. Fig. \ref{fig:levels}). When applying feedback to the system for these parameters, the antibunching can be enhanced, i.e. the stationary value for the second order correlation function is closer to zero (Fig. \ref{fig:antibunching}(b)).

\begin{figure}[!htb]
\includegraphics[width=0.5\columnwidth]{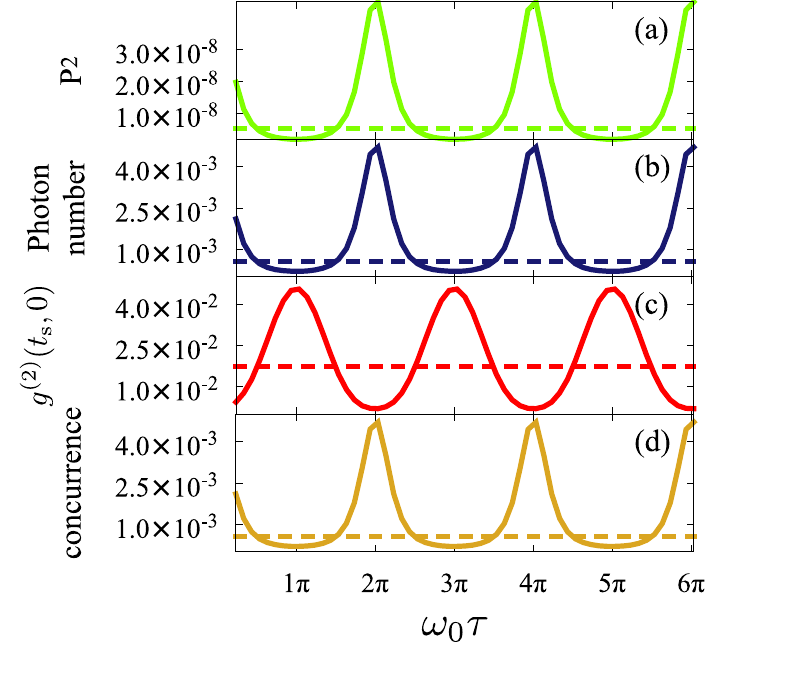}~~~%
	\caption{Impact of short feedback (solid) on the system in the case of antibunching (detuning $\Delta=\sqrt{2}g$). We consider (a) the two photon probability, (b) the photon number, (c) the second order correlation function and (d) the concurrence. All values oscillate due to the alternating constructive and destructive interference of the cavity field with its former output. As reference, the case without feedback is shown as dashed line. The parameters are taken as $\omega_0=1.1\times10^5\gamma$, $g=40\gamma$, $\epsilon=0.035\gamma$, and $G_0=\sqrt{\frac{2c\gamma}{\pi}}$.\label{fig:AntibunchingTau}}
\end{figure}

So far, we discussed the case where the antibunching is enhanced by applying feedback to the system. To achieve this, feedback has to be adjusted to the correct time delay. We now consider the stationary states of the system as a function of the time delay. Considering Fig. \ref{fig:AntibunchingTau}, we observe a periodicity of $\omega_0 \tau$. This corresponds to the change of the phase with which the field interferes with itself. Two important cases will be discussed. Firstly the case of constructive interference, $\omega_0\tau=n 2\pi$, $n=1,2,...$, and secondly, the case of destructive interference $\omega_0\tau=n 2\pi+\pi$, $n=0,1,2,...$.

\begin{figure}[h]
\includegraphics[width=0.35\columnwidth]{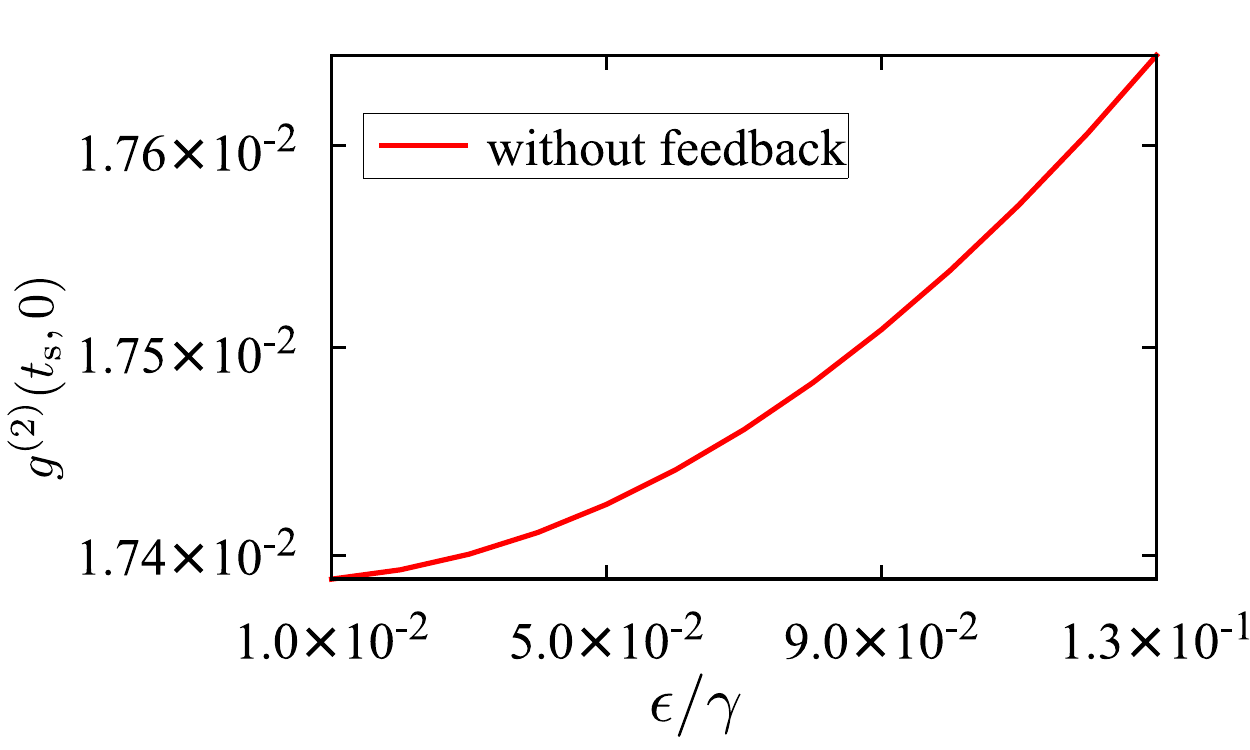}
\caption{ \label{fig:g2drive2ScientificReport} Behavior of the second order correlation function $g^{(2)}(t_{\mathrm{s}},0)$ in the stationary state $t_{\mathrm{s}}\rightarrow\infty$ without feedback for increasing pump strength: By increasing the drive, the antibunching is decreased. Here, the parameters are chosen as $\omega_0=1.1\times10^5\gamma$, and $g=40\gamma$.}
\end{figure}

Now we consider the expectation value for two photon processes, $P_2=\langle\varphi(t)|a^{\dagger} a^{\dagger} aa|\varphi(t)\rangle=2|Cgg20|^2$, which is proportional to the two photon probability in our approximation. This value is given in Fig. \ref{fig:AntibunchingTau}(a). We compare it with the photon number, cf. Fig. \ref{fig:AntibunchingTau}(b), and observe that their maxima occur for the same $\tau$. The antibunching is also maximal when both values are maximal. This is the case when the light field interferes constructively ($\omega_0\tau=n 2\pi$) with itself. This cannot be achieved by simply increasing the intensity in the cavity mode without applying feedback by pumping more strongly, since this decreases the antibunching, as shown in Fig. \ref{fig:g2drive2ScientificReport}. In contrast to this, by applying feedback with constructive interference, the single excitation state is reinforced, increasing antibunching, while at the same time increasing the intensity.

By adjusting the feedback to the destructive interference ($\omega_0\tau=n 2\pi+\pi$), the antibunching is decreased.

Furthermore, we observe that the entanglement is also increased by applying 
feedback. 
This is also achieved for the case of constructive interference. 
The coincidence of the maxima of entanglement with the minima in the second 
order correlation function (maximal antibunching) is consistent with the 
observations in the case without feedback in Ref. \cite{ScieRe}.
Being a strongly dissipative system, this kind of entanglement serves 
merely as an indicator for the non-classical state the combined
emitter-photon system is steered into.
The degree of entanglement is too small to serve as a platform for 
quantum information computing. For these applications schemes exist that allow to create much higher entanglement \cite{RevModPhys.81.865,PhysRevLett.91.177901}
We stress, however, that our platform might be used in setups where
the photon statistics serves as an important feature, e.g. in 
quantum cascaded setups \cite{PhysRevA.94.063825} and quantum optical spectroscopy 
\cite{PhysRevA.73.013813,*PhysRevA.82.013820,*PhysRevB.79.035316}, quantum light pumped induced lasing \cite{PhysRevLett.115.027401},
or for the creation of states with tailored higher order statistics \cite{munoz_emitters_2014}.

\subsection*{Enhancement of bunching}

\begin{figure}[h]
\includegraphics[width=0.5\columnwidth]{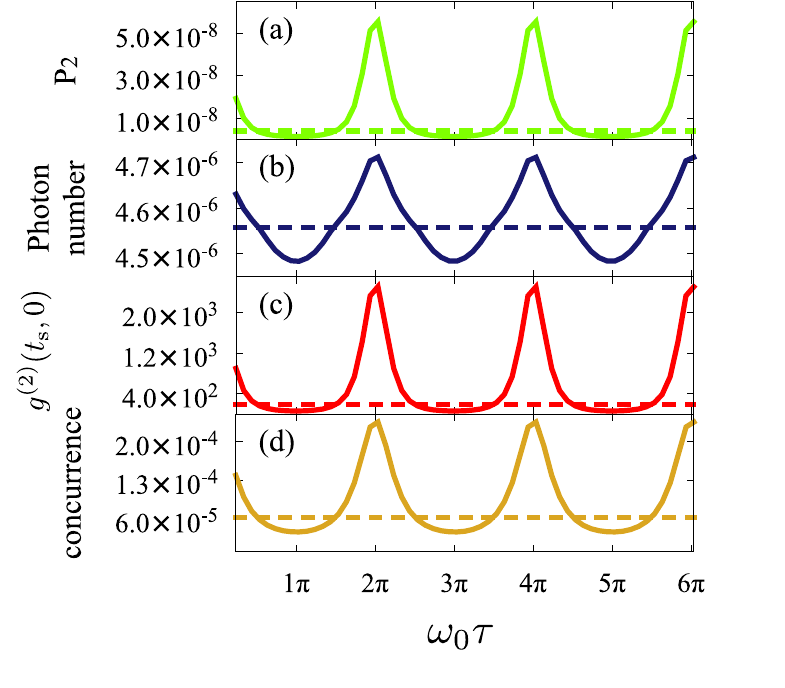}~~~%
	\caption{Impact of short feedback (solid) on the system in the case of bunching (detuning $\Delta=\sqrt{1.5}g$). We consider (a) the two photon probability, (b) the photon number, (c) the second order correlation function and (d) the concurrence. All values oscillate due to the alternating constructive and destructive interference of the cavity field with its former output. As reference, the case without feedback is shown as dashed line. The parameters are taken as $\omega_0=1.1\times10^5\gamma$, $g=40\gamma$, $\epsilon=0.035\gamma$, and $G_0=\sqrt{\frac{2c\gamma}{\pi}}$.\label{fig:bunchingTau}}
\end{figure}

Second, we consider the situation where the non-feedback case exhibits bunching. This is achieved by choosing the detuning $\Delta=\sqrt{1.5} g$, where we address the two photon state $|2_+\rangle$ by a two photon process. When applying feedback to this system, cf. Fig. \ref{fig:bunchingTau}, now the bunching can be enhanced.

Here, again the stationary states with respect to the feedback time is shown in Fig. \ref{fig:bunchingTau}. In this case, for constructive interference ($\omega_0\tau=n 2\pi$), we observe the maximal bunching (cf. Fig. \ref{fig:bunchingTau}(c)). While, again, the photon number as well as the two photon probability are enhanced in this case. Here, however, the enhancement of the two photon probability is more significant. In addition, the entanglement between the two emitters given by the concurrence may also be enhanced by adjusting the feedback time.

Thus, we show that superbunching can be enhanced via quantum feedback.
Superradiance, a many emitter effect, has been shown in such cQED systems
recently and a phase selective coupling leads to strong bunching signatures
\cite{jahnke_giant_2016}.
Our strong bunching enhancement relies not on the number of emitters, but 
on the quantum feedback induced intereference effect.
We achieve an order of magnitude larger bunching in contrast to the non-feedback
case, although the number of photons is increased.
The quantum feedback setup circumvents therefore the typical trade-off between
high number of photons/high number of emitters or strong bunching.

\section{Entanglement and photon statistics}
\label{sec:entanglement}

In this part, we will briefly discuss the relationship between entanglement and photon statistics, which was introduced in Ref. \cite{ScieRe}. In the two scenarios considered in this manuscript, the photon statistics are selected by pumping certain transitions and thus addressing specific states. By analyzing the states addressed by the pumping frequency further, we can observe certain properties.
When choosing the detuning $\Delta=\sqrt{2}g$, where the second order correlation function shows antibunching, the $|1_+\rangle$-state is addressed. Then, the dominant part of the state is still the ground state due to the small pumping rate chosen in our study.
Nonetheless, it is very unlikely to observe more than one photon coincidentally, since the $|1_+\rangle$-state only contains maximally a single photon. This is imprinted onto the output field, as antibunching. This is connected to the entanglement between the two emitters in this system, since this state contains the Bell-state
\begin{equation}
\psi_+=\frac{1}{\sqrt{2}}\left( |g,e\rangle_e  +|e,g\rangle_e \right)
\end{equation}
in the subspace of the emitters indicated by the index $e$, cf. App \ref{sec:states}.
Thus, the anitbunching is connected to entanglement in the above form. By applying feedback to the system, not only the one photon state is reinforced, but indeed the whole entangled state, as can be seen by the simultaneous increase of antibunching and concurrence, which would be not given in the same manner by a simple increase of the pumping strength, cf. Sec. \ref{sec:control}.

This is analogously true for the case of bunching, i.e. $\Delta=\sqrt{1.5}g$, while here the $|2_+\rangle$-state is addressed, which contains two photons, but a more complex type of entanglement. Nonetheless, feedback again reinforces the correlated state. Thus, the relation between photon statistics and entanglement still holds, even when applying feedback on the system.

\section{Conclusion}

In summary, we presented an approximative model for computing time delayed coherent feedback in the Schrödinger picture. By employing this model, we were able to show that time delayed feedback may be used to control the photon statistics of the light field emitted from a cavity. We were interested especially in the cases where the light field shows antibunching and bunching. By adjusting the feedback, these properties may be either enhanced or decreased. Thus, we are able to significantly increase the nonclassical properties of the emitted light field in the case of photon antibunching. This behavior cannot be explained by a simple increase in the field strength, as an analogous increase in the pump strength does not lead to stronger antibunching. Furthermore, the entanglement of the two emitters inside the cavity may be controlled and enhanced in comparison to the non-feedback case.

\begin{acknowledgments}
Y.L. and Q.Z. acknowledge funding by the National Science Foundation (NSF) of China with the Grant No.11675014. Additional support was provided by the Ministry of Science and Technology of China (2013YQ030595-3). This work was supported by International Graduate Exchange Program of Beijing Institute of Technology.
N.L.N., A.K., and A.C. are grateful towards the Deutsche Forschungsgemeinschaft for support
through SFB 910 ‘‘Control of self-organizing nonlinear systems’’ (project B1).
N.L.N also acknowledges support through the School of Nanophotonics (DFG SFB 787).
Y.L. thanks Sven Moritz Hein for very helpful discussions.

\end{acknowledgments}

Y.L. and N.L.N. contributed equally to this work.

\appendix

\section{Equations of motion and correlation function}
\label{sec:eom}

The equations of motion for the time evolution of the state \eqref{eq:psi} are given by the Schr\"{o}dinger equation
\begin{equation}
i\frac{\partial}{\partial t} |\varphi(t)\rangle = H |\varphi(t)\rangle.
\end{equation}
The coefficient $|C_{gg00}|$ is approximated as unity since weak driving is considered, so that the equations of motion become
\begin{widetext}
\begin{align}
\partial_t C_{ge00}&=-i\left(\delta C_{ge00}+g_2C_{gg10}+\epsilon C_{ge10}\right)-\frac{\gamma}{2}C_{ge00}\nonumber\\
\partial_t C_{eg00}&=-i\left(\delta C_{eg00}+g_1C_{gg10}+\epsilon C_{eg10}\right)-\frac{\gamma}{2}C_{eg00}\nonumber\\
\partial_t C_{ee00}&=-i\left(2\delta C_{ee00}+g_1C_{ge10}+g_2C_{eg10}\right)-\gamma C_{ee00}\nonumber\\
\partial_t C_{gg20}&=-i\left(2\Delta C_{gg20}+\sqrt{2}g_2C_{ge10}+\sqrt{2}g_1 C_{eg10}+\sqrt{2}\epsilon C_{gg10}\right)+i\int dk G(k,t)\sqrt{2}C_{gg1k}\nonumber\\
\partial_t C_{ge10}&=-i\left[[\Delta C_{ge10}+\left(\delta-\frac{i\gamma}{2}\right) C_{ge10}+\sqrt{2}g_2C_{gg20}+g_1 C_{ee00}+\epsilon C_{ge00}\right]+i\int dk G(k,t)C_{ge0k}\nonumber\\
\partial_t C_{eg10}&=-i\left[\Delta C_{eg10}+\left(\delta-\frac{i\gamma}{2}\right) C_{eg10}+g_2C_{ee00}+\sqrt{2}g_1 C_{gg20}+\epsilon C_{eg00}\right]+i\int dk G(k,t)C_{eg0k}\nonumber\\
\partial_t C_{ge0k}&=-i\left[\left(\delta-\frac{i\gamma}{2}\right) C_{ge0k}+g_2C_{gg1k}\right]+i G^{*}(k,t)C_{ge10}\nonumber\\
\partial_t C_{eg0k}&=-i\left[\left(\delta-\frac{i\gamma}{2}\right) C_{eg0k}+g_1C_{gg1k}\right]+i G^{*}(k,t)C_{eg10}\nonumber\\
\partial_t C_{gg0k}&=-i\epsilon C_{gg1k}+i G^{*}(k,t)C_{gg10}\nonumber\\
\partial_t C_{gg0kk'}&=i G^{*}(k',t)C_{gg1k}+i G^{*}(k,t)C_{gg1k'}, when \ k\neq k'\nonumber\\
\partial_t C_{gg0kk'}&=i G^{*}(k',t)\sqrt{2}C_{gg1k}, when \ k=k'.\nonumber\\
\partial_t C_{gg1k}&=-i\left[\Delta C_{gg1k}+g_2C_{ge0k}+g_1C_{eg0k}+\epsilon C_{gg0k}\right]\nonumber\\\displaybreak
&+i\left[\int^{k^-}_{-\infty}dpG(p,t)C_{gg0pk}+\int^{+\infty}_{k+}dpG(p,t)C_{ggokp}+dkG(k,t)C_{gg0kk}\sqrt{2}+G^*(k,t)C_{gg20}\sqrt{2}\right]\nonumber\\
\partial_tC_{gg10}&=i\int dk G(k,t)C_{gg0k}-i\left[\Delta C_{gg10}+g_2C_{ge00}+g_1C_{eg00}+\epsilon\left(C_{gg00}+\sqrt{2}C_{gg20}\right)\right].
\label{eq:Sp}
\end{align}
\end{widetext}
The state Eq. \eqref{eq:psi} can be used to evaluate the second order correlation function Eq. \eqref{g2S} explicitly as
\begin{widetext}
\begin{equation}
g^{(2)}(t,0)=\frac{2|C_{gg20}|^2}{(|C_{gg10}|^2+2|C_{gg20}|^2+|C_{ge10}|^2+|C_{eg10}|^2+\int dk |C_{gg1k}|^2)^2}.
\end{equation}
\end{widetext}

\section{States of the system with cavity-emitter coupling}
\label{sec:states}

Here, we give the states used in Sec. \ref{sec:control} to discuss the qualitative behavior of the system. By diagonalizing the the Hamiltonian without feedback reservoir in the above approximation, we get the states \cite{ScieRe}
\begin{align*}
|1_0\rangle&=\frac{1}{\sqrt{2}}|0,g,e\rangle-\frac{1}{\sqrt{2}}|0,e,g\rangle\\
|1_+\rangle&=\frac{1}{\sqrt{2}} |1,g,g\rangle+ \frac{1}{2}|0,g,e\rangle+\frac{1}{2}|0,e,g\rangle\\
|1_-\rangle&=\frac{1}{\sqrt{2}} |1,g,g\rangle- \frac{1}{2}|0,g,e\rangle-\frac{1}{2}|0,e,g\rangle\\
|2_0^1\rangle&=\frac{1}{\sqrt{3}}|2,g,g\rangle-\frac{\sqrt{6}}{3}|0,e,e\rangle\\
|2_0^2\rangle&=\frac{1}{\sqrt{2}}|1,g,e\rangle-\frac{1}{\sqrt{2}}|1,e,g\rangle\\
|2_+\rangle&=\frac{\sqrt{3}}{3}|2,g,g\rangle+\frac{1}{2}|1,g,e\rangle+\frac{1}{2}|1,e,g\rangle+\frac{1}{6}|0,e,e\rangle\\
|2_+\rangle&=\frac{\sqrt{3}}{3}|2,g,g\rangle-\frac{1}{2}|1,g,e\rangle-\frac{1}{2}|1,e,g\rangle+\frac{1}{6}|0,e,e\rangle.
\end{align*}

\bibliographystyle{apsrev4-1}

\begin{thebibliography}{55}%
\makeatletter
\providecommand \@ifxundefined [1]{%
 \@ifx{#1\undefined}
}%
\providecommand \@ifnum [1]{%
 \ifnum #1\expandafter \@firstoftwo
 \else \expandafter \@secondoftwo
 \fi
}%
\providecommand \@ifx [1]{%
 \ifx #1\expandafter \@firstoftwo
 \else \expandafter \@secondoftwo
 \fi
}%
\providecommand \natexlab [1]{#1}%
\providecommand \enquote  [1]{``#1''}%
\providecommand \bibnamefont  [1]{#1}%
\providecommand \bibfnamefont [1]{#1}%
\providecommand \citenamefont [1]{#1}%
\providecommand \href@noop [0]{\@secondoftwo}%
\providecommand \href [0]{\begingroup \@sanitize@url \@href}%
\providecommand \@href[1]{\@@startlink{#1}\@@href}%
\providecommand \@@href[1]{\endgroup#1\@@endlink}%
\providecommand \@sanitize@url [0]{\catcode `\\12\catcode `\$12\catcode
  `\&12\catcode `\#12\catcode `\^12\catcode `\_12\catcode `\%12\relax}%
\providecommand \@@startlink[1]{}%
\providecommand \@@endlink[0]{}%
\providecommand \url  [0]{\begingroup\@sanitize@url \@url }%
\providecommand \@url [1]{\endgroup\@href {#1}{\urlprefix }}%
\providecommand \urlprefix  [0]{URL }%
\providecommand \Eprint [0]{\href }%
\providecommand \doibase [0]{http://dx.doi.org/}%
\providecommand \selectlanguage [0]{\@gobble}%
\providecommand \bibinfo  [0]{\@secondoftwo}%
\providecommand \bibfield  [0]{\@secondoftwo}%
\providecommand \translation [1]{[#1]}%
\providecommand \BibitemOpen [0]{}%
\providecommand \bibitemStop [0]{}%
\providecommand \bibitemNoStop [0]{.\EOS\space}%
\providecommand \EOS [0]{\spacefactor3000\relax}%
\providecommand \BibitemShut  [1]{\csname bibitem#1\endcsname}%
\let\auto@bib@innerbib\@empty
\bibitem [{\citenamefont {P.Zoller}\ \emph {et~al.}(2005)\citenamefont
  {P.Zoller}, \citenamefont {T.Beth}, \citenamefont {D.Binosi}, \citenamefont
  {R.Blatt},\ and\ \citenamefont {H.Briegel}}]{epjd1}%
  \BibitemOpen
  \bibfield  {author} {\bibinfo {author} {\bibnamefont {P.Zoller}}, \bibinfo
  {author} {\bibnamefont {T.Beth}}, \bibinfo {author} {\bibnamefont
  {D.Binosi}}, \bibinfo {author} {\bibnamefont {R.Blatt}}, \ and\ \bibinfo
  {author} {\bibnamefont {H.Briegel}},\ }\href@noop {} {\bibfield  {journal}
  {\bibinfo  {journal} {Eur. Phys. J. D}\ }\textbf {\bibinfo {volume} {36}},\
  \bibinfo {pages} {203} (\bibinfo {year} {2005})}\BibitemShut {NoStop}%
\bibitem [{\citenamefont {Wiseman}\ and\ \citenamefont
  {Milburn}(2006)}]{wiseman1}%
  \BibitemOpen
  \bibfield  {author} {\bibinfo {author} {\bibfnamefont {H.}~\bibnamefont
  {Wiseman}}\ and\ \bibinfo {author} {\bibfnamefont {G.}~\bibnamefont
  {Milburn}},\ }\href@noop {} {\emph {\bibinfo {title} {Quantum Measurement and
  Control}}}\ (\bibinfo  {publisher} {Cambridge University Press, Oxford},\
  \bibinfo {year} {2006})\BibitemShut {NoStop}%
\bibitem [{\citenamefont {van Enk}\ \emph {et~al.}(1997)\citenamefont {van
  Enk}, \citenamefont {Cirac},\ and\ \citenamefont {{P. Zoller}}}]{prl2}%
  \BibitemOpen
  \bibfield  {author} {\bibinfo {author} {\bibfnamefont {S.~J.}\ \bibnamefont
  {van Enk}}, \bibinfo {author} {\bibfnamefont {J.~I.}\ \bibnamefont {Cirac}},
  \ and\ \bibinfo {author} {\bibnamefont {{P. Zoller}}},\ }\href@noop {}
  {\bibfield  {journal} {\bibinfo  {journal} {Phys. Rev. Lett.}\ }\textbf
  {\bibinfo {volume} {79}},\ \bibinfo {pages} {5178} (\bibinfo {year}
  {1997})}\BibitemShut {NoStop}%
\bibitem [{\citenamefont {{Peter W. Shor}}(1995)}]{pra3}%
  \BibitemOpen
  \bibfield  {author} {\bibinfo {author} {\bibnamefont {{Peter W. Shor}}},\
  }\href@noop {} {\bibfield  {journal} {\bibinfo  {journal} {Phys. Rev. A}\
  }\textbf {\bibinfo {volume} {52}},\ \bibinfo {pages} {R2493(R)} (\bibinfo
  {year} {1995})}\BibitemShut {NoStop}%
\bibitem [{\citenamefont {Zhou}\ \emph {et~al.}(2012)\citenamefont {Zhou},
  \citenamefont {Dotsenko}, \citenamefont {Peaudecerf}, \citenamefont
  {Rybarczyk}, \citenamefont {Sayrin}, \citenamefont {Gleyzes}, \citenamefont
  {Raimond}, \citenamefont {Brune},\ and\ \citenamefont {{S. Haroche}}}]{prl4}%
  \BibitemOpen
  \bibfield  {author} {\bibinfo {author} {\bibfnamefont {X.}~\bibnamefont
  {Zhou}}, \bibinfo {author} {\bibfnamefont {I.}~\bibnamefont {Dotsenko}},
  \bibinfo {author} {\bibfnamefont {B.}~\bibnamefont {Peaudecerf}}, \bibinfo
  {author} {\bibfnamefont {T.}~\bibnamefont {Rybarczyk}}, \bibinfo {author}
  {\bibfnamefont {C.}~\bibnamefont {Sayrin}}, \bibinfo {author} {\bibfnamefont
  {S.}~\bibnamefont {Gleyzes}}, \bibinfo {author} {\bibfnamefont {J.~M.}\
  \bibnamefont {Raimond}}, \bibinfo {author} {\bibfnamefont {M.}~\bibnamefont
  {Brune}}, \ and\ \bibinfo {author} {\bibnamefont {{S. Haroche}}},\
  }\href@noop {} {\bibfield  {journal} {\bibinfo  {journal} {Phys. Rev. Lett.}\
  }\textbf {\bibinfo {volume} {108}},\ \bibinfo {pages} {243602} (\bibinfo
  {year} {2012})}\BibitemShut {NoStop}%
\bibitem [{\citenamefont {K.Pyragas}(1992)}]{pla5}%
  \BibitemOpen
  \bibfield  {author} {\bibinfo {author} {\bibnamefont {K.Pyragas}},\
  }\href@noop {} {\bibfield  {journal} {\bibinfo  {journal} {Physics Letters
  A}\ }\textbf {\bibinfo {volume} {170}},\ \bibinfo {pages} {421} (\bibinfo
  {year} {1992})}\BibitemShut {NoStop}%
\bibitem [{\citenamefont {Sch{\"o}ll}\ \emph {et~al.}(2016)\citenamefont
  {Sch{\"o}ll}, \citenamefont {Klapp},\ and\ \citenamefont
  {H{\"o}vel}}]{sfb910}%
  \BibitemOpen
  \bibinfo {editor} {\bibfnamefont {E.}~\bibnamefont {Sch{\"o}ll}}, \bibinfo
  {editor} {\bibfnamefont {S.~H.~L.}\ \bibnamefont {Klapp}}, \ and\ \bibinfo
  {editor} {\bibfnamefont {P.}~\bibnamefont {H{\"o}vel}},\ eds.,\ \href@noop {}
  {\emph {\bibinfo {title} {Control of Self-Organizing Nonlinear Systems}}}\
  (\bibinfo  {publisher} {Springer International Publishing},\ \bibinfo {year}
  {2016})\BibitemShut {NoStop}%
\bibitem [{\citenamefont {Albert}\ \emph
  {et~al.}(2011{\natexlab{a}})\citenamefont {Albert}, \citenamefont {Hopfmann},
  \citenamefont {Reitzenstein}, \citenamefont {Schneider}, \citenamefont
  {H{\"o}fling}, \citenamefont {Worschech}, \citenamefont {Kamp}, \citenamefont
  {Kinzel}, \citenamefont {Forchel},\ and\ \citenamefont {{Ido Kanter}}}]{nc6}%
  \BibitemOpen
  \bibfield  {author} {\bibinfo {author} {\bibfnamefont {F.}~\bibnamefont
  {Albert}}, \bibinfo {author} {\bibfnamefont {C.}~\bibnamefont {Hopfmann}},
  \bibinfo {author} {\bibfnamefont {S.}~\bibnamefont {Reitzenstein}}, \bibinfo
  {author} {\bibfnamefont {C.}~\bibnamefont {Schneider}}, \bibinfo {author}
  {\bibfnamefont {S.}~\bibnamefont {H{\"o}fling}}, \bibinfo {author}
  {\bibfnamefont {L.}~\bibnamefont {Worschech}}, \bibinfo {author}
  {\bibfnamefont {M.}~\bibnamefont {Kamp}}, \bibinfo {author} {\bibfnamefont
  {W.}~\bibnamefont {Kinzel}}, \bibinfo {author} {\bibfnamefont
  {A.}~\bibnamefont {Forchel}}, \ and\ \bibinfo {author} {\bibnamefont {{Ido
  Kanter}}},\ }\href@noop {} {\bibfield  {journal} {\bibinfo  {journal} {Nat.
  Commun.}\ }\textbf {\bibinfo {volume} {2}},\ \bibinfo {pages} {366} (\bibinfo
  {year} {2011}{\natexlab{a}})}\BibitemShut {NoStop}%
\bibitem [{\citenamefont {Hopfmann}\ \emph {et~al.}(2013)\citenamefont
  {Hopfmann}, \citenamefont {Albert}, \citenamefont {Schneider}, \citenamefont
  {Höfling}, \citenamefont {Kamp}, \citenamefont {Forchel}, \citenamefont
  {Kanter},\ and\ \citenamefont {Reitzenstein}}]{1367-2630-15-2-025030}%
  \BibitemOpen
  \bibfield  {author} {\bibinfo {author} {\bibfnamefont {C.}~\bibnamefont
  {Hopfmann}}, \bibinfo {author} {\bibfnamefont {F.}~\bibnamefont {Albert}},
  \bibinfo {author} {\bibfnamefont {C.}~\bibnamefont {Schneider}}, \bibinfo
  {author} {\bibfnamefont {S.}~\bibnamefont {Höfling}}, \bibinfo {author}
  {\bibfnamefont {M.}~\bibnamefont {Kamp}}, \bibinfo {author} {\bibfnamefont
  {A.}~\bibnamefont {Forchel}}, \bibinfo {author} {\bibfnamefont
  {I.}~\bibnamefont {Kanter}}, \ and\ \bibinfo {author} {\bibfnamefont
  {S.}~\bibnamefont {Reitzenstein}},\ }\href
  {http://stacks.iop.org/1367-2630/15/i=2/a=025030} {\bibfield  {journal}
  {\bibinfo  {journal} {New Journal of Physics}\ }\textbf {\bibinfo {volume}
  {15}},\ \bibinfo {pages} {025030} (\bibinfo {year} {2013})}\BibitemShut
  {NoStop}%
\bibitem [{\citenamefont {Hoi}\ \emph {et~al.}(2015)\citenamefont {Hoi},
  \citenamefont {Kockum}, \citenamefont {Tornberg}, \citenamefont
  {Pourkabirian}, \citenamefont {Johansson}, \citenamefont {Delsing},\ and\
  \citenamefont {Wilson}}]{hoi_probing_2015}%
  \BibitemOpen
  \bibfield  {author} {\bibinfo {author} {\bibfnamefont {I.-C.}\ \bibnamefont
  {Hoi}}, \bibinfo {author} {\bibfnamefont {A.~F.}\ \bibnamefont {Kockum}},
  \bibinfo {author} {\bibfnamefont {L.}~\bibnamefont {Tornberg}}, \bibinfo
  {author} {\bibfnamefont {A.}~\bibnamefont {Pourkabirian}}, \bibinfo {author}
  {\bibfnamefont {G.}~\bibnamefont {Johansson}}, \bibinfo {author}
  {\bibfnamefont {P.}~\bibnamefont {Delsing}}, \ and\ \bibinfo {author}
  {\bibfnamefont {C.~M.}\ \bibnamefont {Wilson}},\ }\href
  {http://dx.doi.org/10.1038/nphys3484} {\bibfield  {journal} {\bibinfo
  {journal} {Nat Phys}\ }\textbf {\bibinfo {volume} {11}},\ \bibinfo {pages}
  {1045} (\bibinfo {year} {2015})}\BibitemShut {NoStop}%
\bibitem [{\citenamefont {Lang}\ and\ \citenamefont {Kobayashi}(1980)}]{jqe6}%
  \BibitemOpen
  \bibfield  {author} {\bibinfo {author} {\bibfnamefont {R.}~\bibnamefont
  {Lang}}\ and\ \bibinfo {author} {\bibfnamefont {K.}~\bibnamefont
  {Kobayashi}},\ }\href@noop {} {\bibfield  {journal} {\bibinfo  {journal}
  {IEEE Journal of Quantum Electronics}\ }\textbf {\bibinfo {volume} {QE-16}},\
  \bibinfo {pages} {347} (\bibinfo {year} {1980})}\BibitemShut {NoStop}%
\bibitem [{\citenamefont {Gauthier}\ \emph {et~al.}(1994)\citenamefont
  {Gauthier}, \citenamefont {Sukow}, \citenamefont {Concannon},\ and\
  \citenamefont {{Joshua E. S. Socolar}}}]{pre7}%
  \BibitemOpen
  \bibfield  {author} {\bibinfo {author} {\bibfnamefont {D.~J.}\ \bibnamefont
  {Gauthier}}, \bibinfo {author} {\bibfnamefont {D.~W.}\ \bibnamefont {Sukow}},
  \bibinfo {author} {\bibfnamefont {H.~M.}\ \bibnamefont {Concannon}}, \ and\
  \bibinfo {author} {\bibnamefont {{Joshua E. S. Socolar}}},\ }\href@noop {}
  {\bibfield  {journal} {\bibinfo  {journal} {Phys. Rev. E}\ }\textbf {\bibinfo
  {volume} {50}},\ \bibinfo {pages} {2343} (\bibinfo {year}
  {1994})}\BibitemShut {NoStop}%
\bibitem [{\citenamefont {Otto}\ \emph {et~al.}(2012)\citenamefont {Otto},
  \citenamefont {L{\"u}dge}, \citenamefont {Vladimirov}, \citenamefont
  {Wolfrum},\ and\ \citenamefont {{E Sch{\"o}ll}}}]{njp8}%
  \BibitemOpen
  \bibfield  {author} {\bibinfo {author} {\bibfnamefont {C.}~\bibnamefont
  {Otto}}, \bibinfo {author} {\bibfnamefont {K.}~\bibnamefont {L{\"u}dge}},
  \bibinfo {author} {\bibfnamefont {A.~G.}\ \bibnamefont {Vladimirov}},
  \bibinfo {author} {\bibfnamefont {M.}~\bibnamefont {Wolfrum}}, \ and\
  \bibinfo {author} {\bibnamefont {{E Sch{\"o}ll}}},\ }\href@noop {} {\bibfield
   {journal} {\bibinfo  {journal} {New J. Phys.}\ }\textbf {\bibinfo {volume}
  {14}},\ \bibinfo {pages} {113033} (\bibinfo {year} {2012})}\BibitemShut
  {NoStop}%
\bibitem [{\citenamefont {Munnelly}\ \emph {et~al.}(2017)\citenamefont
  {Munnelly}, \citenamefont {Lingnau}, \citenamefont {Karow}, \citenamefont
  {Heindel}, \citenamefont {Kamp}, \citenamefont {H\"ofling}, \citenamefont
  {L\"udge}, \citenamefont {Schneider},\ and\ \citenamefont
  {Reitzenstein}}]{2017_03_Munnelly}%
  \BibitemOpen
  \bibfield  {author} {\bibinfo {author} {\bibfnamefont {P.}~\bibnamefont
  {Munnelly}}, \bibinfo {author} {\bibfnamefont {B.}~\bibnamefont {Lingnau}},
  \bibinfo {author} {\bibfnamefont {M.~M.}\ \bibnamefont {Karow}}, \bibinfo
  {author} {\bibfnamefont {T.}~\bibnamefont {Heindel}}, \bibinfo {author}
  {\bibfnamefont {M.}~\bibnamefont {Kamp}}, \bibinfo {author} {\bibfnamefont
  {S.}~\bibnamefont {H\"ofling}}, \bibinfo {author} {\bibfnamefont
  {K.}~\bibnamefont {L\"udge}}, \bibinfo {author} {\bibfnamefont
  {C.}~\bibnamefont {Schneider}}, \ and\ \bibinfo {author} {\bibfnamefont
  {S.}~\bibnamefont {Reitzenstein}},\ }\href {\doibase 10.1364/OPTICA.4.000303}
  {\bibfield  {journal} {\bibinfo  {journal} {Optica}\ }\textbf {\bibinfo
  {volume} {4}},\ \bibinfo {pages} {303} (\bibinfo {year} {2017})}\BibitemShut
  {NoStop}%
\bibitem [{\citenamefont {Sudhir}\ \emph {et~al.}(2017)\citenamefont {Sudhir},
  \citenamefont {Wilson}, \citenamefont {Schilling}, \citenamefont {Sch\"utz},
  \citenamefont {Fedorov}, \citenamefont {Ghadimi}, \citenamefont
  {Nunnenkamp},\ and\ \citenamefont {Kippenberg}}]{PhysRevX.7.011001}%
  \BibitemOpen
  \bibfield  {author} {\bibinfo {author} {\bibfnamefont {V.}~\bibnamefont
  {Sudhir}}, \bibinfo {author} {\bibfnamefont {D.~J.}\ \bibnamefont {Wilson}},
  \bibinfo {author} {\bibfnamefont {R.}~\bibnamefont {Schilling}}, \bibinfo
  {author} {\bibfnamefont {H.}~\bibnamefont {Sch\"utz}}, \bibinfo {author}
  {\bibfnamefont {S.~A.}\ \bibnamefont {Fedorov}}, \bibinfo {author}
  {\bibfnamefont {A.~H.}\ \bibnamefont {Ghadimi}}, \bibinfo {author}
  {\bibfnamefont {A.}~\bibnamefont {Nunnenkamp}}, \ and\ \bibinfo {author}
  {\bibfnamefont {T.~J.}\ \bibnamefont {Kippenberg}},\ }\href {\doibase
  10.1103/PhysRevX.7.011001} {\bibfield  {journal} {\bibinfo  {journal} {Phys.
  Rev. X}\ }\textbf {\bibinfo {volume} {7}},\ \bibinfo {pages} {011001}
  (\bibinfo {year} {2017})}\BibitemShut {NoStop}%
\bibitem [{\citenamefont {Jacobs}\ \emph {et~al.}(2014)\citenamefont {Jacobs},
  \citenamefont {Wang},\ and\ \citenamefont {Wiseman}}]{1367-2630-16-7-073036}%
  \BibitemOpen
  \bibfield  {author} {\bibinfo {author} {\bibfnamefont {K.}~\bibnamefont
  {Jacobs}}, \bibinfo {author} {\bibfnamefont {X.}~\bibnamefont {Wang}}, \ and\
  \bibinfo {author} {\bibfnamefont {H.~M.}\ \bibnamefont {Wiseman}},\ }\href
  {http://stacks.iop.org/1367-2630/16/i=7/a=073036} {\bibfield  {journal}
  {\bibinfo  {journal} {New Journal of Physics}\ }\textbf {\bibinfo {volume}
  {16}},\ \bibinfo {pages} {073036} (\bibinfo {year} {2014})}\BibitemShut
  {NoStop}%
\bibitem [{\citenamefont {Dorner}\ and\ \citenamefont
  {Zoller}(2002)}]{UDornerPZoller}%
  \BibitemOpen
  \bibfield  {author} {\bibinfo {author} {\bibfnamefont {U.}~\bibnamefont
  {Dorner}}\ and\ \bibinfo {author} {\bibfnamefont {P.}~\bibnamefont
  {Zoller}},\ }\href@noop {} {\bibfield  {journal} {\bibinfo  {journal} {Phys.
  Rev. A}\ }\textbf {\bibinfo {volume} {66}},\ \bibinfo {pages} {023816}
  (\bibinfo {year} {2002})}\BibitemShut {NoStop}%
\bibitem [{\citenamefont {Milonni}\ and\ \citenamefont
  {Knight}(1974)}]{PhysRevA.10.1096}%
  \BibitemOpen
  \bibfield  {author} {\bibinfo {author} {\bibfnamefont {P.~W.}\ \bibnamefont
  {Milonni}}\ and\ \bibinfo {author} {\bibfnamefont {P.~L.}\ \bibnamefont
  {Knight}},\ }\href {\doibase 10.1103/PhysRevA.10.1096} {\bibfield  {journal}
  {\bibinfo  {journal} {Phys. Rev. A}\ }\textbf {\bibinfo {volume} {10}},\
  \bibinfo {pages} {1096} (\bibinfo {year} {1974})}\BibitemShut {NoStop}%
\bibitem [{\citenamefont {Breuer}\ \emph {et~al.}(2016)\citenamefont {Breuer},
  \citenamefont {Laine}, \citenamefont {Piilo},\ and\ \citenamefont
  {Vacchini}}]{RevModPhys.88.021002}%
  \BibitemOpen
  \bibfield  {author} {\bibinfo {author} {\bibfnamefont {H.-P.}\ \bibnamefont
  {Breuer}}, \bibinfo {author} {\bibfnamefont {E.-M.}\ \bibnamefont {Laine}},
  \bibinfo {author} {\bibfnamefont {J.}~\bibnamefont {Piilo}}, \ and\ \bibinfo
  {author} {\bibfnamefont {B.}~\bibnamefont {Vacchini}},\ }\href@noop {}
  {\bibfield  {journal} {\bibinfo  {journal} {Rev. Mod. Phys.}\ }\textbf
  {\bibinfo {volume} {88}},\ \bibinfo {pages} {021002} (\bibinfo {year}
  {2016})}\BibitemShut {NoStop}%
\bibitem [{\citenamefont {Strunz}\ \emph {et~al.}(1999)\citenamefont {Strunz},
  \citenamefont {Di{\'o}si},\ and\ \citenamefont
  {Gisin}}]{PhysRevLett.82.1801}%
  \BibitemOpen
  \bibfield  {author} {\bibinfo {author} {\bibfnamefont {W.~T.}\ \bibnamefont
  {Strunz}}, \bibinfo {author} {\bibfnamefont {L.}~\bibnamefont {Di{\'o}si}}, \
  and\ \bibinfo {author} {\bibfnamefont {N.}~\bibnamefont {Gisin}},\
  }\href@noop {} {\bibfield  {journal} {\bibinfo  {journal} {Phys. Rev. Lett.}\
  }\textbf {\bibinfo {volume} {82}},\ \bibinfo {pages} {1801} (\bibinfo {year}
  {1999})}\BibitemShut {NoStop}%
\bibitem [{\citenamefont {Rivas}\ \emph {et~al.}(2014)\citenamefont {Rivas},
  \citenamefont {Huelga},\ and\ \citenamefont
  {Plenio}}]{0034-4885-77-9-094001}%
  \BibitemOpen
  \bibfield  {author} {\bibinfo {author} {\bibfnamefont {{\'A}.}~\bibnamefont
  {Rivas}}, \bibinfo {author} {\bibfnamefont {S.~F.}\ \bibnamefont {Huelga}}, \
  and\ \bibinfo {author} {\bibfnamefont {M.~B.}\ \bibnamefont {Plenio}},\
  }\href@noop {} {\bibfield  {journal} {\bibinfo  {journal} {Reports on
  Progress in Physics}\ }\textbf {\bibinfo {volume} {77}},\ \bibinfo {pages}
  {094001} (\bibinfo {year} {2014})}\BibitemShut {NoStop}%
\bibitem [{\citenamefont {Naumann}\ \emph {et~al.}(2014)\citenamefont
  {Naumann}, \citenamefont {Hein}, \citenamefont {Knorr},\ and\ \citenamefont
  {{Julia Kabuss}}}]{pra11}%
  \BibitemOpen
  \bibfield  {author} {\bibinfo {author} {\bibfnamefont {N.~L.}\ \bibnamefont
  {Naumann}}, \bibinfo {author} {\bibfnamefont {S.~M.}\ \bibnamefont {Hein}},
  \bibinfo {author} {\bibfnamefont {A.}~\bibnamefont {Knorr}}, \ and\ \bibinfo
  {author} {\bibnamefont {{Julia Kabuss}}},\ }\href@noop {} {\bibfield
  {journal} {\bibinfo  {journal} {Phys. Rev. A}\ }\textbf {\bibinfo {volume}
  {90}},\ \bibinfo {pages} {043835} (\bibinfo {year} {2014})}\BibitemShut
  {NoStop}%
\bibitem [{\citenamefont {Kopylov}\ \emph {et~al.}(2015)\citenamefont
  {Kopylov}, \citenamefont {Radonji{\'c}}, \citenamefont {Brandes},
  \citenamefont {Bala\v{z}},\ and\ \citenamefont {{Axel Pelster}}}]{pra12}%
  \BibitemOpen
  \bibfield  {author} {\bibinfo {author} {\bibfnamefont {W.}~\bibnamefont
  {Kopylov}}, \bibinfo {author} {\bibfnamefont {M.}~\bibnamefont
  {Radonji{\'c}}}, \bibinfo {author} {\bibfnamefont {T.}~\bibnamefont
  {Brandes}}, \bibinfo {author} {\bibfnamefont {A.}~\bibnamefont {Bala\v{z}}},
  \ and\ \bibinfo {author} {\bibnamefont {{Axel Pelster}}},\ }\href@noop {}
  {\bibfield  {journal} {\bibinfo  {journal} {Phys. Rev. A}\ }\textbf {\bibinfo
  {volume} {92}},\ \bibinfo {pages} {063832} (\bibinfo {year}
  {2015})}\BibitemShut {NoStop}%
\bibitem [{\citenamefont {Kabuss}\ \emph {et~al.}(2016)\citenamefont {Kabuss},
  \citenamefont {Katsch}, \citenamefont {Knorr},\ and\ \citenamefont
  {{Alexander Carmele}}}]{josa19}%
  \BibitemOpen
  \bibfield  {author} {\bibinfo {author} {\bibfnamefont {J.}~\bibnamefont
  {Kabuss}}, \bibinfo {author} {\bibfnamefont {F.}~\bibnamefont {Katsch}},
  \bibinfo {author} {\bibfnamefont {A.}~\bibnamefont {Knorr}}, \ and\ \bibinfo
  {author} {\bibnamefont {{Alexander Carmele}}},\ }\href@noop {} {\bibfield
  {journal} {\bibinfo  {journal} {Journal of the Optical Society of America B}\
  }\textbf {\bibinfo {volume} {33(7)}},\ \bibinfo {pages} {C10} (\bibinfo
  {year} {2016})}\BibitemShut {NoStop}%
\bibitem [{\citenamefont {Albert}\ \emph
  {et~al.}(2011{\natexlab{b}})\citenamefont {Albert}, \citenamefont {Hopfmann},
  \citenamefont {Reitzenstein}, \citenamefont {Schneider}, \citenamefont
  {Höfling}, \citenamefont {Worschech}, \citenamefont {Kamp}, \citenamefont
  {Kinzel}, \citenamefont {Forchel},\ and\ \citenamefont
  {Kanter}}]{albert_observing_2011}%
  \BibitemOpen
  \bibfield  {author} {\bibinfo {author} {\bibfnamefont {F.}~\bibnamefont
  {Albert}}, \bibinfo {author} {\bibfnamefont {C.}~\bibnamefont {Hopfmann}},
  \bibinfo {author} {\bibfnamefont {S.}~\bibnamefont {Reitzenstein}}, \bibinfo
  {author} {\bibfnamefont {C.}~\bibnamefont {Schneider}}, \bibinfo {author}
  {\bibfnamefont {S.}~\bibnamefont {Höfling}}, \bibinfo {author}
  {\bibfnamefont {L.}~\bibnamefont {Worschech}}, \bibinfo {author}
  {\bibfnamefont {M.}~\bibnamefont {Kamp}}, \bibinfo {author} {\bibfnamefont
  {W.}~\bibnamefont {Kinzel}}, \bibinfo {author} {\bibfnamefont
  {A.}~\bibnamefont {Forchel}}, \ and\ \bibinfo {author} {\bibfnamefont
  {I.}~\bibnamefont {Kanter}},\ }\href {http://dx.doi.org/10.1038/ncomms1370}
  {\bibfield  {journal} {\bibinfo  {journal} {Nature Communications}\ }\textbf
  {\bibinfo {volume} {2}},\ \bibinfo {pages} {366} (\bibinfo {year}
  {2011}{\natexlab{b}})}\BibitemShut {NoStop}%
\bibitem [{\citenamefont {Carmele}\ \emph {et~al.}(2013)\citenamefont
  {Carmele}, \citenamefont {Kabuss}, \citenamefont {Schulze}, \citenamefont
  {Reitzenstein},\ and\ \citenamefont {{Andreas Knorr}}}]{prl13}%
  \BibitemOpen
  \bibfield  {author} {\bibinfo {author} {\bibfnamefont {A.}~\bibnamefont
  {Carmele}}, \bibinfo {author} {\bibfnamefont {J.}~\bibnamefont {Kabuss}},
  \bibinfo {author} {\bibfnamefont {F.}~\bibnamefont {Schulze}}, \bibinfo
  {author} {\bibfnamefont {S.}~\bibnamefont {Reitzenstein}}, \ and\ \bibinfo
  {author} {\bibnamefont {{Andreas Knorr}}},\ }\href@noop {} {\bibfield
  {journal} {\bibinfo  {journal} {Phys. Rev. Lett.}\ }\textbf {\bibinfo
  {volume} {110}},\ \bibinfo {pages} {013601} (\bibinfo {year}
  {2013})}\BibitemShut {NoStop}%
\bibitem [{\citenamefont {Hein}\ \emph {et~al.}(2014)\citenamefont {Hein},
  \citenamefont {Schulze}, \citenamefont {Carmele},\ and\ \citenamefont
  {{Andreas Knorr}}}]{prl14}%
  \BibitemOpen
  \bibfield  {author} {\bibinfo {author} {\bibfnamefont {S.~M.}\ \bibnamefont
  {Hein}}, \bibinfo {author} {\bibfnamefont {F.}~\bibnamefont {Schulze}},
  \bibinfo {author} {\bibfnamefont {A.}~\bibnamefont {Carmele}}, \ and\
  \bibinfo {author} {\bibnamefont {{Andreas Knorr}}},\ }\href@noop {}
  {\bibfield  {journal} {\bibinfo  {journal} {Phys. Rev. Lett.}\ }\textbf
  {\bibinfo {volume} {113}},\ \bibinfo {pages} {027401} (\bibinfo {year}
  {2014})}\BibitemShut {NoStop}%
\bibitem [{\citenamefont {Hein}\ \emph {et~al.}(2015)\citenamefont {Hein},
  \citenamefont {Schulze}, \citenamefont {Carmele},\ and\ \citenamefont
  {{Andreas Knorr}}}]{pra15}%
  \BibitemOpen
  \bibfield  {author} {\bibinfo {author} {\bibfnamefont {S.~M.}\ \bibnamefont
  {Hein}}, \bibinfo {author} {\bibfnamefont {F.}~\bibnamefont {Schulze}},
  \bibinfo {author} {\bibfnamefont {A.}~\bibnamefont {Carmele}}, \ and\
  \bibinfo {author} {\bibnamefont {{Andreas Knorr}}},\ }\href@noop {}
  {\bibfield  {journal} {\bibinfo  {journal} {Phys. Rev. A}\ }\textbf {\bibinfo
  {volume} {91}},\ \bibinfo {pages} {052321} (\bibinfo {year}
  {2015})}\BibitemShut {NoStop}%
\bibitem [{\citenamefont {Kraft}\ \emph {et~al.}(2016)\citenamefont {Kraft},
  \citenamefont {Hein}, \citenamefont {Lehnert}, \citenamefont {Sch\"oll},
  \citenamefont {Hughes},\ and\ \citenamefont {Knorr}}]{pra16}%
  \BibitemOpen
  \bibfield  {author} {\bibinfo {author} {\bibfnamefont {M.}~\bibnamefont
  {Kraft}}, \bibinfo {author} {\bibfnamefont {S.~M.}\ \bibnamefont {Hein}},
  \bibinfo {author} {\bibfnamefont {J.}~\bibnamefont {Lehnert}}, \bibinfo
  {author} {\bibfnamefont {E.}~\bibnamefont {Sch\"oll}}, \bibinfo {author}
  {\bibfnamefont {S.}~\bibnamefont {Hughes}}, \ and\ \bibinfo {author}
  {\bibfnamefont {A.}~\bibnamefont {Knorr}},\ }\href@noop {} {\bibfield
  {journal} {\bibinfo  {journal} {Phys. Rev. A}\ }\textbf {\bibinfo {volume}
  {94}},\ \bibinfo {pages} {023806} (\bibinfo {year} {2016})}\BibitemShut
  {NoStop}%
\bibitem [{\citenamefont {N{\'e}met}\ and\ \citenamefont
  {Parkins}(2016)}]{PhysRevA.94.023809}%
  \BibitemOpen
  \bibfield  {author} {\bibinfo {author} {\bibfnamefont {N.}~\bibnamefont
  {N{\'e}met}}\ and\ \bibinfo {author} {\bibfnamefont {S.}~\bibnamefont
  {Parkins}},\ }\href@noop {} {\bibfield  {journal} {\bibinfo  {journal} {Phys.
  Rev. A}\ }\textbf {\bibinfo {volume} {94}},\ \bibinfo {pages} {023809}
  (\bibinfo {year} {2016})}\BibitemShut {NoStop}%
\bibitem [{\citenamefont {{Arne L. Grimsmo}}(2015)}]{prl17}%
  \BibitemOpen
  \bibfield  {author} {\bibinfo {author} {\bibnamefont {{Arne L. Grimsmo}}},\
  }\href@noop {} {\bibfield  {journal} {\bibinfo  {journal} {Phys. Rev. Lett.}\
  }\textbf {\bibinfo {volume} {115}},\ \bibinfo {pages} {060402} (\bibinfo
  {year} {2015})}\BibitemShut {NoStop}%
\bibitem [{\citenamefont {Pichler}\ and\ \citenamefont {{Peter
  Zoller}}(2016)}]{prl18}%
  \BibitemOpen
  \bibfield  {author} {\bibinfo {author} {\bibfnamefont {H.}~\bibnamefont
  {Pichler}}\ and\ \bibinfo {author} {\bibnamefont {{Peter Zoller}}},\
  }\href@noop {} {\bibfield  {journal} {\bibinfo  {journal} {Phys. Rev. Lett.}\
  }\textbf {\bibinfo {volume} {116}},\ \bibinfo {pages} {093601} (\bibinfo
  {year} {2016})}\BibitemShut {NoStop}%
\bibitem [{\citenamefont {Zhang}\ \emph {et~al.}(2016)\citenamefont {Zhang},
  \citenamefont {Zhang},\ and\ \citenamefont {{Chang-shui Yu}}}]{ScieRe}%
  \BibitemOpen
  \bibfield  {author} {\bibinfo {author} {\bibfnamefont {Y.}~\bibnamefont
  {Zhang}}, \bibinfo {author} {\bibfnamefont {J.}~\bibnamefont {Zhang}}, \ and\
  \bibinfo {author} {\bibnamefont {{Chang-shui Yu}}},\ }\href@noop {}
  {\bibfield  {journal} {\bibinfo  {journal} {Scientific Reports}\ }\textbf
  {\bibinfo {volume} {6}},\ \bibinfo {pages} {24098} (\bibinfo {year}
  {2016})}\BibitemShut {NoStop}%
\bibitem [{\citenamefont {Gardiner}\ and\ \citenamefont
  {Zoller}(2004)}]{quantumnoise}%
  \BibitemOpen
  \bibfield  {author} {\bibinfo {author} {\bibfnamefont {C.}~\bibnamefont
  {Gardiner}}\ and\ \bibinfo {author} {\bibfnamefont {P.}~\bibnamefont
  {Zoller}},\ }\href@noop {} {\emph {\bibinfo {title} {Quantum Noise}}}\
  (\bibinfo  {publisher} {Springer-Verlag Berlin Heidelberg},\ \bibinfo {year}
  {2004})\BibitemShut {NoStop}%
\bibitem [{\citenamefont {Lu}\ \emph {et~al.}(2015)\citenamefont {Lu},
  \citenamefont {Liu},\ and\ \citenamefont {Zhao}}]{Lu2015161}%
  \BibitemOpen
  \bibfield  {author} {\bibinfo {author} {\bibfnamefont {Y.}~\bibnamefont
  {Lu}}, \bibinfo {author} {\bibfnamefont {H.}~\bibnamefont {Liu}}, \ and\
  \bibinfo {author} {\bibfnamefont {Q.}~\bibnamefont {Zhao}},\ }\href {\doibase
  http://dx.doi.org/10.1016/j.aop.2015.04.030} {\bibfield  {journal} {\bibinfo
  {journal} {Annals of Physics}\ }\textbf {\bibinfo {volume} {360}},\ \bibinfo
  {pages} {161 } (\bibinfo {year} {2015})}\BibitemShut {NoStop}%
\bibitem [{\citenamefont {Lu}\ and\ \citenamefont
  {Zhao}(2016)}]{lu_minimum_2016}%
  \BibitemOpen
  \bibfield  {author} {\bibinfo {author} {\bibfnamefont {Y.}~\bibnamefont
  {Lu}}\ and\ \bibinfo {author} {\bibfnamefont {Q.}~\bibnamefont {Zhao}},\
  }\href {http://dx.doi.org/10.1038/srep32057} {\bibfield  {journal} {\bibinfo
  {journal} {Scientific Reports}\ }\textbf {\bibinfo {volume} {6}},\ \bibinfo
  {pages} {32057} (\bibinfo {year} {2016})}\BibitemShut {NoStop}%
\bibitem [{\citenamefont {James}\ \emph {et~al.}(2001)\citenamefont {James},
  \citenamefont {Kwiat}, \citenamefont {Munro},\ and\ \citenamefont
  {White}}]{PhysRevA.64.052312}%
  \BibitemOpen
  \bibfield  {author} {\bibinfo {author} {\bibfnamefont {D.~F.~V.}\
  \bibnamefont {James}}, \bibinfo {author} {\bibfnamefont {P.~G.}\ \bibnamefont
  {Kwiat}}, \bibinfo {author} {\bibfnamefont {W.~J.}\ \bibnamefont {Munro}}, \
  and\ \bibinfo {author} {\bibfnamefont {A.~G.}\ \bibnamefont {White}},\ }\href
  {\doibase 10.1103/PhysRevA.64.052312} {\bibfield  {journal} {\bibinfo
  {journal} {Phys. Rev. A}\ }\textbf {\bibinfo {volume} {64}},\ \bibinfo
  {pages} {052312} (\bibinfo {year} {2001})}\BibitemShut {NoStop}%
\bibitem [{\citenamefont {Thew}\ \emph {et~al.}(2002)\citenamefont {Thew},
  \citenamefont {Nemoto}, \citenamefont {White},\ and\ \citenamefont
  {Munro}}]{PhysRevA.66.012303}%
  \BibitemOpen
  \bibfield  {author} {\bibinfo {author} {\bibfnamefont {R.~T.}\ \bibnamefont
  {Thew}}, \bibinfo {author} {\bibfnamefont {K.}~\bibnamefont {Nemoto}},
  \bibinfo {author} {\bibfnamefont {A.~G.}\ \bibnamefont {White}}, \ and\
  \bibinfo {author} {\bibfnamefont {W.~J.}\ \bibnamefont {Munro}},\ }\href
  {\doibase 10.1103/PhysRevA.66.012303} {\bibfield  {journal} {\bibinfo
  {journal} {Phys. Rev. A}\ }\textbf {\bibinfo {volume} {66}},\ \bibinfo
  {pages} {012303} (\bibinfo {year} {2002})}\BibitemShut {NoStop}%
\bibitem [{\citenamefont {Sapienza}\ \emph {et~al.}(2015)\citenamefont
  {Sapienza}, \citenamefont {Davanço}, \citenamefont {Badolato},\ and\
  \citenamefont {Srinivasan}}]{sapienza_nanoscale_2015}%
  \BibitemOpen
  \bibfield  {author} {\bibinfo {author} {\bibfnamefont {L.}~\bibnamefont
  {Sapienza}}, \bibinfo {author} {\bibfnamefont {M.}~\bibnamefont {Davanço}},
  \bibinfo {author} {\bibfnamefont {A.}~\bibnamefont {Badolato}}, \ and\
  \bibinfo {author} {\bibfnamefont {K.}~\bibnamefont {Srinivasan}},\ }\href
  {http://dx.doi.org/10.1038/ncomms8833} {\bibfield  {journal} {\bibinfo
  {journal} {Nature Communications}\ }\textbf {\bibinfo {volume} {6}},\
  \bibinfo {pages} {7833} (\bibinfo {year} {2015})}\BibitemShut {NoStop}%
\bibitem [{\citenamefont {Kabuss}\ \emph {et~al.}(2015)\citenamefont {Kabuss},
  \citenamefont {O.Krimer}, \citenamefont {Rotter}, \citenamefont {Stannigel},
  \citenamefont {Knorr},\ and\ \citenamefont {Carmele}}]{JuliaPRA}%
  \BibitemOpen
  \bibfield  {author} {\bibinfo {author} {\bibfnamefont {J.}~\bibnamefont
  {Kabuss}}, \bibinfo {author} {\bibfnamefont {D.}~\bibnamefont {O.Krimer}},
  \bibinfo {author} {\bibfnamefont {S.}~\bibnamefont {Rotter}}, \bibinfo
  {author} {\bibfnamefont {K.}~\bibnamefont {Stannigel}}, \bibinfo {author}
  {\bibfnamefont {A.}~\bibnamefont {Knorr}}, \ and\ \bibinfo {author}
  {\bibfnamefont {A.}~\bibnamefont {Carmele}},\ }\href@noop {} {\bibfield
  {journal} {\bibinfo  {journal} {Phys. Rev. A}\ }\textbf {\bibinfo {volume}
  {92}},\ \bibinfo {pages} {053801} (\bibinfo {year} {2015})}\BibitemShut
  {NoStop}%
\bibitem [{\citenamefont {Walls}\ and\ \citenamefont {Milburn}(1994)}]{Wall}%
  \BibitemOpen
  \bibfield  {author} {\bibinfo {author} {\bibfnamefont {D.}~\bibnamefont
  {Walls}}\ and\ \bibinfo {author} {\bibfnamefont {G.}~\bibnamefont
  {Milburn}},\ }\href@noop {} {\bibfield  {journal} {\bibinfo  {journal}
  {Quantum optics}\ }\textbf {\bibinfo {volume} {3}},\ \bibinfo {pages} {39}
  (\bibinfo {year} {1994})}\BibitemShut {NoStop}%
\bibitem [{\citenamefont {Wootters}(1998)}]{Woot}%
  \BibitemOpen
  \bibfield  {author} {\bibinfo {author} {\bibfnamefont {W.}~\bibnamefont
  {Wootters}},\ }\href@noop {} {\bibfield  {journal} {\bibinfo  {journal}
  {Phys. Rev. Lett.}\ }\textbf {\bibinfo {volume} {80}},\ \bibinfo {pages}
  {2245} (\bibinfo {year} {1998})}\BibitemShut {NoStop}%
\bibitem [{\citenamefont {Sch\"on}\ \emph {et~al.}(2005)\citenamefont
  {Sch\"on}, \citenamefont {Solano}, \citenamefont {Verstraete}, \citenamefont
  {Cirac},\ and\ \citenamefont {Wolf}}]{PhysRevLett.95.110503}%
  \BibitemOpen
  \bibfield  {author} {\bibinfo {author} {\bibfnamefont {C.}~\bibnamefont
  {Sch\"on}}, \bibinfo {author} {\bibfnamefont {E.}~\bibnamefont {Solano}},
  \bibinfo {author} {\bibfnamefont {F.}~\bibnamefont {Verstraete}}, \bibinfo
  {author} {\bibfnamefont {J.~I.}\ \bibnamefont {Cirac}}, \ and\ \bibinfo
  {author} {\bibfnamefont {M.~M.}\ \bibnamefont {Wolf}},\ }\href {\doibase
  10.1103/PhysRevLett.95.110503} {\bibfield  {journal} {\bibinfo  {journal}
  {Phys. Rev. Lett.}\ }\textbf {\bibinfo {volume} {95}},\ \bibinfo {pages}
  {110503} (\bibinfo {year} {2005})}\BibitemShut {NoStop}%
\bibitem [{\citenamefont {Sch\"on}\ \emph {et~al.}(2007)\citenamefont
  {Sch\"on}, \citenamefont {Hammerer}, \citenamefont {Wolf}, \citenamefont
  {Cirac},\ and\ \citenamefont {Solano}}]{PhysRevA.75.032311}%
  \BibitemOpen
  \bibfield  {author} {\bibinfo {author} {\bibfnamefont {C.}~\bibnamefont
  {Sch\"on}}, \bibinfo {author} {\bibfnamefont {K.}~\bibnamefont {Hammerer}},
  \bibinfo {author} {\bibfnamefont {M.~M.}\ \bibnamefont {Wolf}}, \bibinfo
  {author} {\bibfnamefont {J.~I.}\ \bibnamefont {Cirac}}, \ and\ \bibinfo
  {author} {\bibfnamefont {E.}~\bibnamefont {Solano}},\ }\href {\doibase
  10.1103/PhysRevA.75.032311} {\bibfield  {journal} {\bibinfo  {journal} {Phys.
  Rev. A}\ }\textbf {\bibinfo {volume} {75}},\ \bibinfo {pages} {032311}
  (\bibinfo {year} {2007})}\BibitemShut {NoStop}%
\bibitem [{\citenamefont {Schollwöck}(2011)}]{Schollwoeck201196}%
  \BibitemOpen
  \bibfield  {author} {\bibinfo {author} {\bibfnamefont {U.}~\bibnamefont
  {Schollwöck}},\ }\href {\doibase
  http://dx.doi.org/10.1016/j.aop.2010.09.012} {\bibfield  {journal} {\bibinfo
  {journal} {Annals of Physics}\ }\textbf {\bibinfo {volume} {326}},\ \bibinfo
  {pages} {96 } (\bibinfo {year} {2011})},\ \bibinfo {note} {january 2011
  Special Issue}\BibitemShut {NoStop}%
\bibitem [{\citenamefont {Vidal}(2003)}]{PhysRevLett.91.147902}%
  \BibitemOpen
  \bibfield  {author} {\bibinfo {author} {\bibfnamefont {G.}~\bibnamefont
  {Vidal}},\ }\href {\doibase 10.1103/PhysRevLett.91.147902} {\bibfield
  {journal} {\bibinfo  {journal} {Phys. Rev. Lett.}\ }\textbf {\bibinfo
  {volume} {91}},\ \bibinfo {pages} {147902} (\bibinfo {year}
  {2003})}\BibitemShut {NoStop}%
\bibitem [{\citenamefont {Horodecki}\ \emph {et~al.}(2009)\citenamefont
  {Horodecki}, \citenamefont {Horodecki}, \citenamefont {Horodecki},\ and\
  \citenamefont {Horodecki}}]{RevModPhys.81.865}%
  \BibitemOpen
  \bibfield  {author} {\bibinfo {author} {\bibfnamefont {R.}~\bibnamefont
  {Horodecki}}, \bibinfo {author} {\bibfnamefont {P.}~\bibnamefont
  {Horodecki}}, \bibinfo {author} {\bibfnamefont {M.}~\bibnamefont
  {Horodecki}}, \ and\ \bibinfo {author} {\bibfnamefont {K.}~\bibnamefont
  {Horodecki}},\ }\href {\doibase 10.1103/RevModPhys.81.865} {\bibfield
  {journal} {\bibinfo  {journal} {Rev. Mod. Phys.}\ }\textbf {\bibinfo {volume}
  {81}},\ \bibinfo {pages} {865} (\bibinfo {year} {2009})}\BibitemShut
  {NoStop}%
\bibitem [{\citenamefont {Clark}\ \emph {et~al.}(2003)\citenamefont {Clark},
  \citenamefont {Peng}, \citenamefont {Gu},\ and\ \citenamefont
  {Parkins}}]{PhysRevLett.91.177901}%
  \BibitemOpen
  \bibfield  {author} {\bibinfo {author} {\bibfnamefont {S.}~\bibnamefont
  {Clark}}, \bibinfo {author} {\bibfnamefont {A.}~\bibnamefont {Peng}},
  \bibinfo {author} {\bibfnamefont {M.}~\bibnamefont {Gu}}, \ and\ \bibinfo
  {author} {\bibfnamefont {S.}~\bibnamefont {Parkins}},\ }\href {\doibase
  10.1103/PhysRevLett.91.177901} {\bibfield  {journal} {\bibinfo  {journal}
  {Phys. Rev. Lett.}\ }\textbf {\bibinfo {volume} {91}},\ \bibinfo {pages}
  {177901} (\bibinfo {year} {2003})}\BibitemShut {NoStop}%
\bibitem [{\citenamefont {Carre\~no}\ and\ \citenamefont
  {Laussy}(2016)}]{PhysRevA.94.063825}%
  \BibitemOpen
  \bibfield  {author} {\bibinfo {author} {\bibfnamefont {J.~C.~L.}\
  \bibnamefont {Carre\~no}}\ and\ \bibinfo {author} {\bibfnamefont {F.~P.}\
  \bibnamefont {Laussy}},\ }\href {\doibase 10.1103/PhysRevA.94.063825}
  {\bibfield  {journal} {\bibinfo  {journal} {Phys. Rev. A}\ }\textbf {\bibinfo
  {volume} {94}},\ \bibinfo {pages} {063825} (\bibinfo {year}
  {2016})}\BibitemShut {NoStop}%
\bibitem [{\citenamefont {Kira}\ and\ \citenamefont
  {Koch}(2006)}]{PhysRevA.73.013813}%
  \BibitemOpen
  \bibfield  {author} {\bibinfo {author} {\bibfnamefont {M.}~\bibnamefont
  {Kira}}\ and\ \bibinfo {author} {\bibfnamefont {S.~W.}\ \bibnamefont
  {Koch}},\ }\href {\doibase 10.1103/PhysRevA.73.013813} {\bibfield  {journal}
  {\bibinfo  {journal} {Phys. Rev. A}\ }\textbf {\bibinfo {volume} {73}},\
  \bibinfo {pages} {013813} (\bibinfo {year} {2006})}\BibitemShut {NoStop}%
\bibitem [{\citenamefont {Richter}\ and\ \citenamefont
  {Mukamel}(2010)}]{PhysRevA.82.013820}%
  \BibitemOpen
  \bibfield  {author} {\bibinfo {author} {\bibfnamefont {M.}~\bibnamefont
  {Richter}}\ and\ \bibinfo {author} {\bibfnamefont {S.}~\bibnamefont
  {Mukamel}},\ }\href {\doibase 10.1103/PhysRevA.82.013820} {\bibfield
  {journal} {\bibinfo  {journal} {Phys. Rev. A}\ }\textbf {\bibinfo {volume}
  {82}},\ \bibinfo {pages} {013820} (\bibinfo {year} {2010})}\BibitemShut
  {NoStop}%
\bibitem [{\citenamefont {Carmele}\ \emph {et~al.}(2009)\citenamefont
  {Carmele}, \citenamefont {Knorr},\ and\ \citenamefont
  {Richter}}]{PhysRevB.79.035316}%
  \BibitemOpen
  \bibfield  {author} {\bibinfo {author} {\bibfnamefont {A.}~\bibnamefont
  {Carmele}}, \bibinfo {author} {\bibfnamefont {A.}~\bibnamefont {Knorr}}, \
  and\ \bibinfo {author} {\bibfnamefont {M.}~\bibnamefont {Richter}},\ }\href
  {\doibase 10.1103/PhysRevB.79.035316} {\bibfield  {journal} {\bibinfo
  {journal} {Phys. Rev. B}\ }\textbf {\bibinfo {volume} {79}},\ \bibinfo
  {pages} {035316} (\bibinfo {year} {2009})}\BibitemShut {NoStop}%
\bibitem [{\citenamefont {Kazimierczuk}\ \emph {et~al.}(2015)\citenamefont
  {Kazimierczuk}, \citenamefont {Schmutzler}, \citenamefont {A\ss{}mann},
  \citenamefont {Schneider}, \citenamefont {Kamp}, \citenamefont {H\"ofling},\
  and\ \citenamefont {Bayer}}]{PhysRevLett.115.027401}%
  \BibitemOpen
  \bibfield  {author} {\bibinfo {author} {\bibfnamefont {T.}~\bibnamefont
  {Kazimierczuk}}, \bibinfo {author} {\bibfnamefont {J.}~\bibnamefont
  {Schmutzler}}, \bibinfo {author} {\bibfnamefont {M.}~\bibnamefont
  {A\ss{}mann}}, \bibinfo {author} {\bibfnamefont {C.}~\bibnamefont
  {Schneider}}, \bibinfo {author} {\bibfnamefont {M.}~\bibnamefont {Kamp}},
  \bibinfo {author} {\bibfnamefont {S.}~\bibnamefont {H\"ofling}}, \ and\
  \bibinfo {author} {\bibfnamefont {M.}~\bibnamefont {Bayer}},\ }\href
  {\doibase 10.1103/PhysRevLett.115.027401} {\bibfield  {journal} {\bibinfo
  {journal} {Phys. Rev. Lett.}\ }\textbf {\bibinfo {volume} {115}},\ \bibinfo
  {pages} {027401} (\bibinfo {year} {2015})}\BibitemShut {NoStop}%
\bibitem [{\citenamefont {Munoz}\ \emph {et~al.}(2014)\citenamefont {Munoz},
  \citenamefont {del Valle}, \citenamefont {Tudela}, \citenamefont
  {{MullerK.}}, \citenamefont {{LichtmanneckerS.}}, \citenamefont
  {{KaniberM.}}, \citenamefont {{TejedorC.}}, \citenamefont {{FinleyJ. J.}},\
  and\ \citenamefont {{LaussyF. P.}}}]{munoz_emitters_2014}%
  \BibitemOpen
  \bibfield  {author} {\bibinfo {author} {\bibfnamefont {C.~S.}\ \bibnamefont
  {Munoz}}, \bibinfo {author} {\bibfnamefont {E.}~\bibnamefont {del Valle}},
  \bibinfo {author} {\bibfnamefont {A.~G.}\ \bibnamefont {Tudela}}, \bibinfo
  {author} {\bibnamefont {{MullerK.}}}, \bibinfo {author} {\bibnamefont
  {{LichtmanneckerS.}}}, \bibinfo {author} {\bibnamefont {{KaniberM.}}},
  \bibinfo {author} {\bibnamefont {{TejedorC.}}}, \bibinfo {author}
  {\bibnamefont {{FinleyJ. J.}}}, \ and\ \bibinfo {author} {\bibnamefont
  {{LaussyF. P.}}},\ }\href {http://dx.doi.org/10.1038/nphoton.2014.114}
  {\bibfield  {journal} {\bibinfo  {journal} {Nat Photon}\ }\textbf {\bibinfo
  {volume} {8}},\ \bibinfo {pages} {550} (\bibinfo {year} {2014})}\BibitemShut
  {NoStop}%
\bibitem [{\citenamefont {Jahnke}\ \emph {et~al.}(2016)\citenamefont {Jahnke},
  \citenamefont {Gies}, \citenamefont {Aßmann}, \citenamefont {Bayer},
  \citenamefont {Leymann}, \citenamefont {Foerster}, \citenamefont {Wiersig},
  \citenamefont {Schneider}, \citenamefont {Kamp},\ and\ \citenamefont
  {Höfling}}]{jahnke_giant_2016}%
  \BibitemOpen
  \bibfield  {author} {\bibinfo {author} {\bibfnamefont {F.}~\bibnamefont
  {Jahnke}}, \bibinfo {author} {\bibfnamefont {C.}~\bibnamefont {Gies}},
  \bibinfo {author} {\bibfnamefont {M.}~\bibnamefont {Aßmann}}, \bibinfo
  {author} {\bibfnamefont {M.}~\bibnamefont {Bayer}}, \bibinfo {author}
  {\bibfnamefont {H.~A.~M.}\ \bibnamefont {Leymann}}, \bibinfo {author}
  {\bibfnamefont {A.}~\bibnamefont {Foerster}}, \bibinfo {author}
  {\bibfnamefont {J.}~\bibnamefont {Wiersig}}, \bibinfo {author} {\bibfnamefont
  {C.}~\bibnamefont {Schneider}}, \bibinfo {author} {\bibfnamefont
  {M.}~\bibnamefont {Kamp}}, \ and\ \bibinfo {author} {\bibfnamefont
  {S.}~\bibnamefont {Höfling}},\ }\href
  {http://dx.doi.org/10.1038/ncomms11540} {\bibfield  {journal} {\bibinfo
  {journal} {Nature Communications}\ }\textbf {\bibinfo {volume} {7}},\
  \bibinfo {pages} {11540} (\bibinfo {year} {2016})}\BibitemShut {NoStop}%
\end{thebibliography}

%

\end{document}